\pdfoutput=1
\documentclass{article}

\usepackage{microtype}
\usepackage[english]{babel}
\usepackage{authblk}
\usepackage{subcaption}
\usepackage{afterpage}
\usepackage{float}
\usepackage{multirow}
\usepackage{amssymb, amsmath, mathtools, tikz}
\usetikzlibrary{arrows}
\usepackage[english]{babel}
\usepackage{amsthm}
\usepackage{algorithm}
\usepackage{algpseudocode}
\usepackage{tikz}
\usepackage{comment}
\usepackage{booktabs}
\usetikzlibrary{arrows}
\usepackage{graphicx}
\usepackage{enumerate}
\usepackage{enumitem}
\usepackage{csquotes}
\usepackage[style=authoryear, giveninits=false, style=apa,backend=biber, doi=false]{biblatex}
\DeclareFieldFormat[article]{title}{``#1''}  

\makeatletter
\DeclareRobustCommand{\MakeSentenceCase}{%
  \@ifstar{\MakeSentenceCaseStar}{\MakeSentenceCaseStar}%
}
\newcommand{\MakeSentenceCaseStar}[1]{#1} 
\makeatother

\usepackage[hidelinks]{hyperref}
\usepackage[letterpaper,top=2cm,bottom=2cm,left=3cm,right=3cm,marginparwidth=1.75cm]{geometry}

\usepackage{amsthm}
\usepackage{amsmath}
\usepackage{amssymb}
\usepackage{graphicx}
\usepackage{xcolor}

\newtheorem{proposition}{Proposition}

\title{Network Modeling of Asynchronous Change-Points in Multivariate Time Series}

\author{
    Carson McKee$^\dagger$ and Maria Kalli \\
    Department of Mathematics, King's College London \\
    $^\dagger$Correspondence: carson.mckee@kcl.ac.uk
}

\date{}
\newcommand{\probP}{\text{I\kern-0.15em P}}

\newcommand{\bX}{\boldsymbol{X}}
\newcommand{\bx}{\boldsymbol{x}}

\newcommand{\bxp}{\boldsymbol{x}_{t-1}}
\newcommand{\bA}{\boldsymbol{A}}
\newcommand{\bW}{\boldsymbol{W}}

\begin{document}
\maketitle

\begin{abstract}
This article introduces a novel Bayesian method for asynchronous change-point detection in multivariate time series. This method allows for change-points to occur earlier in some (leading) series followed, after a short delay, by change-points in some other (lagging) series. Such dynamic dependence structure is common in fields such as seismology and neurology where a latent event such as an earthquake or seizure causes certain sensors to register change-points before others. We model these lead-lag dependencies via a latent directed graph and provide a hierarchical prior for learning the graph's structure and parameters. Posterior inference is made tractable by modifying particle MCMC methods designed for univariate change-point problems. We apply our method to both simulated and real datasets from the fields of seismology and neurology. In the simulated data, we find that our method outperforms competing methods in settings where the change-point locations are dependent across series. In the real data applications we show that our model can also uncover interpretable network structure. 

\par
\vspace{1em}

\end{abstract}

\section{Introduction}
Change-point models are used to infer the locations of abrupt changes in the distribution of a time series. These models assume that a potentially unknown number of change-points partition the data into a set of segments. The data within each segment are assumed to be generated from a common model, which varies from one segment to another. Examples include Gaussian models with changing means and variances \parencite[]{Chernoff1964, Inclan1994} or regression curves with discontinuous jumps \parencite[]{Fearnhead2005}. These models are applied to a range of problems, including the detection of epileptic seizures \parencite[]{Schroder2019}, the identification of economic regime shifts \parencite[]{Maheu2018}, and the detection of earthquakes \parencite[]{Xie2019}. 
\par 
In this article, we present a novel Bayesian method for detecting change-points in multivariate time series.
Our method allows for dependence between pairs of series such that the change-points in some (lagging) series are likely to occur shortly after those in other (leading) series. Identifying such lead-lag relationships is important both for prediction, since detecting a change in a leading series may signal an imminent change in a lagging series, and for inference, by providing insight into the latent processes connecting the series. We represent each series as a node on a latent directed graph whereby a directed edge from one series/node to another indicates the presence of a lead-lag relationship, the strength of which is controlled by a set of edge parameters.
\par 
Although lead-lag relationships arise in many fields, this work is motivated by two specific examples. The first concerns the detection of seismic events using observations from multiple ground motion sensors. In this setting, an event such as an earthquake induces abrupt changes in the amplitude and frequency content of the sensor signals as the primary (P) and secondary (S) waves arrive. Sensors located closer to the fault line typically register change-points earlier than those further away, giving rise to a lead-lag relationship across the multiple series. Here, the lead-lag relationships are driven by spatial proximity, and modelling them may improve detection accuracy of weaker events at more distant sensors.
\par 
The other motivating example is the detection of seizures using electroencephalogram (EEG) signals. EEG analysis involves recording the brain's electrical activity over time using electrodes placed on the scalp. The occurrence of a seizure results in sudden changes in the amplitude and rhythmic patterns of the EEG signals. These changes are initially detected in the electrodes closest to the seizure onset zone and typically propagate to nearby electrodes shortly thereafter, as the seizure spreads to other regions of the brain. In this context, lead-lag relationships are driven by the seizure's dynamics and propagation pathway. Inferring this pathway is crucial for understanding the brain's network structure and for developing personalized treatment strategies aimed at preventing the seizure from spreading.
\par

Early work on change-point detection took a frequentist approach. Univariate methods can be traced back to \textcite{Page1954} who proposed the Cumulative Sum (CUSUM) test for detecting a single change-point in the mean. Multiple change-points can be handled by applying such tests recursively using algorithms such as Binary Segmentation (BS) \parencite[]{Vos81} or Wild Binary Segmentation (WBS) \parencite[]{Fryzlewicz2014}, a stochastic extension to BS which is more sensitive to multiple changes. Another popular approach is to minimize a cost function which assesses the fit of a particular segmentation relative to the number of change-points. Recent work in this area includes the Pruned Exact Linear Time (PELT) method \parencite[]{Killick2012} which performs this minimization in linear time. Reviews of various methods in this area can be found in \textcite{Horvath2014, Truong2020}.
\par 
In the multivariate setting, many frequentist approaches have focused on situations where change-points affect subsets of the series at synchronized times \parencite[see for example][]{Jeng2013, Wang2018} and are therefore not appropriate for the problem at hand. Two exceptions to this are the methods of \textcite{Xie2019} and \textcite{Fisch2022} who allow for change-points to occur with either a fixed delay or within some time window of those in other series. However, assuming a fixed delay may be overly restrictive, and these methods do not account for the lead-lag dependencies considered in this paper.
\par
Bayesian approaches require a prior on the change-points and provide posteriors that quantify uncertainty about their number and locations. In the univariate setting, the early work by \textcite{Chernoff1964} and \textcite{Yao1984} proposed using a geometric renewal process, whereby the random arrival times determine the change locations. \textcite{Barry1992, Barry1993} extended this by using Product Partition Models (PPMs) \parencite[]{Hartigan1990}. PPMs assume that the prior over the induced partition can be written as a product of segment-wise cohesion functions, which includes renewal processes as a special case (when the cohesions depend only on the segment length). Alternative approaches based on Markov chains have also been proposed. For instance, \textcite{Chib1998} models the change-points using a finite-state, unidirectional Markov chain and compares models with different numbers of change-points via Bayes factors. An overview of Bayesian methods and their computational aspects can be found in \textcite{Eckley}. 
\par
When considering multivariate time series, one could apply existing univariate Bayesian methods directly. For example, we may assume a single change-point prior which affects all series simultaneously \parencite[]{Corradin2022} or assign independent change-point priors for different parameters of the model \parencite[]{Peluso2019, Pedroso2023}. However, these approaches are overly restrictive, assuming either complete dependence or independence in the change locations. 
\par 
An alternative and more flexible approach is to allow each time point in each series to have its own change-point probability. Then induce dependence by assigning a joint prior over the vector of probabilities for each time. This was the approach of \textcite{Harle2016} who specified a Dirichlet distribution over all possible change-point configurations at each time. However, the number of parameters in the prior grows exponentially with the number of series, quickly becoming infeasible. An alternative, empirical Bayesian approach was proposed by \textcite{Fan2017} who assumed that the change-point probabilities are shared across series for each fixed time but may vary across time. This prior therefore encourages synchronized change-points across series. 
\par
Another fully Bayesian approach, inspired by the univariate PPM \parencite[]{Barry1992}, is the method of \textcite{Quinlan2024}. Here, the vector of change probabilities at each time are generated by taking the logistic transform of random variables sampled from a multivariate Student's t-distribution. Thus, the correlation between the change-point probabilities at each fixed time can be controlled through the covariance matrix. A related problem was considered by \textcite{Bardwell2017} who proposed the Bayesian Abnormal Segments Detector (BARD) method. This method models the change-point locations with a Markov chain and attempts to identify segments where a subset of the series differ from some baseline behavior. Although, this approach differs from standard change-point methods in that it assumes the series will return to some baseline behavior. 
\par 
The multivariate approaches discussed so far rely on subsets of series changing simultaneously, making them unsuitable for the lead-lag dependencies considered here. A Bayesian method that allows for asynchronous changes was proposed by \textcite{Hallgren2024}, who assumed that the series are connected via a known undirected graph. Under this method, series which are connected on the graph are encouraged to experience changes at similar times. However, because the graph is undirected, this method cannot capture directed lead–lag relationships. It also requires a prespecified graph, which may not be known a priori.
\par 
Our approach to modeling multivariate change-points is similar to that of \textcite{Hallgren2024}. We also assume that the series are connected via a latent weighted graph. However, unlike their approach, we do not assume that the graph is known a priori, and we provide a hierarchical prior to enable learning on the graph parameters. Since we are interested in lead-lag relationships, we use a directed graph to account for directed dependence between pairs of series. We show how posterior inference can be performed by adapting the efficient particle Gibbs sampler of \textcite{Whiteley2011} to the multivariate setting.
\par 
The remainder of this article is organized as follows: Section \ref{sec:model} introduces the proposed model and discusses its properties. In Section \ref{sec:post}, we describe a novel blocked particle Gibbs scheme to obtain posterior samples. In Section \ref{sec:illus}, we demonstrate our method through a comparative simulation study and two real data examples before concluding with a discussion in Section \ref{sec:disc}. Proofs of results and background information on the particle Gibbs sampler are given in the Appendices. 
All code and data required to reproduce the plots, tables and analysis is available at \href{https://github.com/carsonmckee/NetCP}{\texttt{github.com/carsonmckee/NetCP}}.

\section{The Model} \label{sec:model}
In this section, we introduce our modeling approach, which we refer to as the Network Change-Point (NetCP) model, and related notation. We consider a multivariate time series of dimension $d$, observed at times $t=1,..., T$. Let $\boldsymbol{Y}_t = (Y_{1,t}, ..., Y_{d, t})^{'}$ denote the vector of observations at time $t$ and $\boldsymbol{Y}=(\boldsymbol{Y}_1, ..., \boldsymbol{Y}_T)$ be the $(d \times T)$ matrix which holds the observations. Conditional on a set of change-point locations, the data in each series is partitioned into non-overlapping segments such that the change-points correspond to the segment end-points. We use the notation $\boldsymbol{Y}_{j, s:t} = (Y_{j, s+1}, ..., Y_{j, t})$ to denote the segment of observations in series $j$ between change-points $s$ and $t$. 
\par 
We model the change-point locations with a multivariate hidden state process, which we refer to as the NetCP process. Each observation $Y_{j, t}$ is assigned a hidden state $X_{j, t}$, which denotes the time since the last change-point before time $t$ in series $j$. That is, $X_{j, t}$ is the \emph{run length} at time $t$ in series $j$. Therefore, at time $t$, the run length $X_{j, t}$ either resets to 1 (indicating a change-point) or increases by 1 (no change-point). Our prior over the hidden states is parameterised according to a latent directed graph, with edges linking the pairs of series that are dependent. We focus on posterior inference for both the hidden states and the graph parameters. 

\subsection{Likelihood Model} \label{sec:like_model}
We first describe the likelihood model conditional on the hidden state process. Recall that $X_{j, t}$ denotes the run length at time $t$ in series $j$. Conditional on $\bX = \bx$, the set of change-point locations in series $j$ is given by
\begin{equation*}
    \boldsymbol{\tau}_{j} = \{i \in \{0, ..., T-1\}:x_{j, i+1} = 1\} \cup \{T\}
\end{equation*}
The change-points partition the observations in each series $j$ into $|\boldsymbol{\tau}_j|-1$ segments, where the $k$th segment is given by $\tau_{j, k}:\tau_{j, k+1} = \{\tau_{j, k} + 1, ..., \tau_{j, k+1}\}$. We assume that the observations in the $k$th segment of the $j$th series have density $f_j(\boldsymbol{y}_{j, \tau_{j, k}:\tau_{j, k+1}} | \theta_{j, k})$ where the parameter $\theta_{j, k} \in \Theta$ is segment specific and drawn independently from the density $f_j(\cdot)$. 
\par 
For computational reasons, \parencite[see][]{Barry1993, Fearnhead2006}, the densities $f_j(\cdot | \theta)$ and $f_j(\cdot)$ are chosen to be conjugate, allowing for the analytic marginalization over the segment specific parameters, i.e.
\begin{equation}\label{eq:marg_like}
    f_j(\boldsymbol{y}_{j, s:t}) = \int_{\Theta} f_j(\boldsymbol{y}_{j, s:t} | \theta) f_j(\theta) d\theta
\end{equation}
where the density $f_j(\boldsymbol{y}_{j, s:t})$ is referred to as the marginal segment density. Throughout this paper, we assume that Equation (\ref{eq:marg_like}) is available analytically or can be approximated numerically in low dimensions. Under these assumptions, the joint likelihood can be expressed as
\begin{eqnarray} \label{eq:joint_like}
    f(\boldsymbol{y}|\bx) &=& \prod_{j=1}^d f_j(\boldsymbol{y}_{j, 0:T}|\bx_{j, 0:T}) \\
    &=& \prod_{j=1}^d \prod_{t=1}^{T}f_j(y_{j, t}|x_{j, t}, \boldsymbol{y}_{j, 0:t-1})
\end{eqnarray} 
where 
\begin{equation} \label{eq:pred_like}
    f_j(y_{j, t}|x_{j, t}, \boldsymbol{y}_{j, 0:t-1}) = \frac{f_j(\boldsymbol{y}_{j, (t-x_{j, t}) : t})}{f_j(\boldsymbol{y}_{j, (t-x_{j, t}) : t-1})}
\end{equation}
can be computed from the marginal segment densities given in Equation (\ref{eq:marg_like}). The likelihood structure given in Equations (\ref{eq:marg_like}) - (\ref{eq:pred_like}) has been widely used in both the univariate and multivariate change-point literature \parencite[]{Barry1993, Fan2017, Hallgren2024}.
\par 
The choice of $f_j(\cdot| \theta)$ and $f_j(\cdot)$ is problem-specific. In Section \ref{sec:illus} we apply our model to seismology and EEG data, where the focus is to detect changes in the variability and/or autocorrelation structure of signals with constant mean. To do this, we use an independent autoregressive model for the data within each segment of each series, such that both the lag coefficients and variance can vary across segments. In particular, we assume that the segment-specific parameters are $\theta := (\boldsymbol{\phi}, \sigma^2)$ where $\boldsymbol{\phi} \in \mathbb{R}^L$ is the vector of autoregressive parameters and $L$ is the number of lags. For a given segment $\boldsymbol{y}_{j, s:t} = \{y_{j, s+1}, ..., y_{j, t}\}$, the likelihood is given by
\begin{equation}\label{eq:ar_likelihood}
    f_j(\boldsymbol{y}_{j, s:t}|\boldsymbol{\phi}, \sigma^2) = \mathcal{N}_{t-s}(\boldsymbol{y}_{j, s:t} | \boldsymbol{H}_{j, st} \boldsymbol{\phi}, \sigma^2\boldsymbol{I}_{t-s})
\end{equation}
where $\mathcal{N}_{t-s}(\cdot|\boldsymbol{\mu}, \boldsymbol{\Sigma})$ denotes the density of the $t-s$ dimensional normal distribution, $\boldsymbol{I}_{t-s}$ denotes the $(t-s)\times (t-s)$ identity matrix and
\begin{equation}
    \boldsymbol{H}_{j, st} = \begin{bmatrix}
        y_{j, s} & y_{j, s-1} & \cdots & y_{j, s-L+1}\\
        y_{j, s+1} & y_{j, s} & \cdots & y_{j, s-L+2} \\
        \vdots & \vdots & \ddots & \vdots \\
        y_{j, t-1} & y_{j, t-2} & \cdots & y_{j, t-L}
    \end{bmatrix}
\end{equation}
is a $(t-s) \times L$ matrix with rows containing the lagged values for each observation in the segment. We place a conjugate normal-inverse-gamma prior over $\theta=(\boldsymbol{\phi}, \sigma^2)$ within each series:
\begin{eqnarray}
    f_j(\phi_{l}|\sigma^2) &=& \mathcal{N}(\phi_{l}|0, \delta_{j,l}\,\sigma^2) \;\; \text{for } l=1, ..., L \\
    f_j(\sigma^2) &=& \mathcal{IG}(\sigma^2 | \alpha_j, \beta_j)
\end{eqnarray}
where $\mathcal{IG}(\sigma^2|\alpha, \beta)$ denotes the density of the inverse gamma distribution and $(\alpha_j, \beta_j, \delta_{j, 1}, ..., \delta_{j,L})$ are the set of hyperparameters for series $j$. Integrating over $(\boldsymbol{\phi}, \sigma^2)$ within a given segment, we obtain the marginal segment density:
\begin{equation} \label{eq:ar_model}
    f_j(\boldsymbol{y}_{j, s:t}) = (2\pi)^{-(t-s)/2} \left(\frac{|\boldsymbol{D}_{j, st}|}{|\boldsymbol{D}_j|}\right)^{1/2} \frac{\beta_j^{\alpha_j}\Gamma(\alpha_{j, st})}{\beta_{j, st}^{\alpha_{j, st}}\Gamma(\alpha_j)}
\end{equation}
where $\Gamma(\cdot)$ denotes the gamma function, $\boldsymbol{D}_j = \text{diag}(\delta_{j, 1}, ..., \delta_{j, L})$, $\alpha_{j, st} = \alpha_j + (t-s)/2$,  $\beta_{j, st} = \beta_{j} + \frac{1}{2}[\|\boldsymbol{Y}_{j, s:t}\|_2^2 - \boldsymbol{E}_{j, st}\boldsymbol{D}_{j, st}\boldsymbol{E}_{j, st}^{'}]$, $\boldsymbol{E}_{j, st} = \boldsymbol{Y}_{j, s:t}\boldsymbol{H}_{j, st}$ and $\boldsymbol{D}_{j, st} = (\boldsymbol{H}_{j, st}^{'}\boldsymbol{H}_{j, st} + \boldsymbol{D}_j^{-1})^{-1}$. 
This likelihood model was previously considered by \textcite{Fearnhead2007} in the univariate setting and is used to model data with a constant mean and change-points in the variability and/or autocorrelation structure. 
\par 
To minimize the error rate in the change-point locations, it is essential to carefully select the hyperparameters $(\alpha_j, \beta_j, \delta_{j,1}, \ldots, \delta_{j,L})$ for each series. A fully Bayesian approach involves placing priors on the hyperparameters and updating them during posterior sampling. However, fixing the parameters can yield substantial computational savings, as each marginal segment density must only be computed once and can be reused. In the examples presented in Section \ref{sec:illus}, we present sensible default values.

\subsection{Change-Point Prior} \label{sec:cp_model}
We now describe our prior on the hidden state process, which we refer to as the NetCP process. The change-point locations in each series are determined by the multivariate hidden state process $\{\bX_t\}_{t=1, ..., T}$ where $\bX_t = (X_{1, t}, ..., X_{d, t})^{'}$ is the vector of run lengths for each series. Since the first segment in each series starts at time $t=1$, the process is initialized by 
\begin{equation} \label{eq:initial_state}
    X_{j, 1}  \overset{a.s.}{=} 1 \;\; \forall \;\; j=1, ..., d
\end{equation}
and for $t=2, ..., T$ we have $\bX_t \in \{1, ..., t\}^d$, with the event $X_{j, t} = 1$ indicating a change-point occurred at time $t-1$ in series $j$. Our prior over the hidden states assumes that the process $\{\bX_t\}_{t=1, ..., T}$ follows a Markov chain initialized by Equation (\ref{eq:initial_state}) with transition probabilities given by
\begin{equation} \label{eq:transition}
    Pr(\bX_t=\bx_t|\bX_{t-1} =\bx_{t-1}) = \prod_{j=1}^d \Pr(X_{j, t} = x_{j, t}|\bX_{t-1}=\bx_{t-1})
\end{equation}
where
\begin{equation} \label{eq:transition_j}
    \Pr(X_{j, t} = x_{j, t}|\bX_{t-1}=\bx_{t-1}) = \begin{cases}
        p_{j, t}(\bx_{t-1}), \;\;\;\;\;\;\;\;\;\;\;\; \text{if   } x_{j, t} = 1 \\
        1-p_{j, t}(\bx_{t-1}), \;\;\;\;\;\;\text{if   } x_{j, t} = x_{j, t-1} + 1
    \end{cases}
\end{equation}
Conditional on $\bX_{t-1}=\bx_{t-1}$, the prior probability that a change-point occurs at time $t-1$ in series $j$ is given by $p_{j, t}(\bx_{t-1})$. The dependence in the change-point locations across the series is controlled by our choice of $p_{j, t}(\bx_{t-1})$. 
\par
A natural starting point in specifying $p_{j, t}(\bx_{t-1})$ is to consider independent univariate change-point priors for each series $j$. A popular choice is the Bernoulli process \parencite[see][]{Yao1984, Barry1993}, where
\begin{equation} \label{eq:bernoulli}
    p_{j, t}(\bx_{t-1}) = q_{0, j} \;\;\; \text{for } j=1, ..., d \;\;\; \text{and  } q_{0, j} \in (0, 1),
\end{equation}
implying that the change-point probabilities in each series are constant for all $t$. Here, the change-point locations in each series $j$ form a renewal process with geometrically distributed inter-arrival times where $q_{0, j}$ is the distribution parameter. Similar to the approach of \textcite{Hallgren2024}, we use the Bernoulli process as a starting point such that, in the case of no dependence, our prior defaults to Equation (\ref{eq:bernoulli}).
\par 
Now, in order to incorporate lead-lag dependencies, if change-points in series $i$ lead those in series $j$, then $p_{j,t}(\bx_{t-1})$ should depend on $x_{i,t-1}$. That is, the probability of a change-point occurring in series $j$ should depend on the distance (run length) to the most recent change-point in series $i$. We represent this dependence via a weighted directed graph, such that the presence of an edge $(i, j)$ indicates that change-points in $j$ are likely to follow change-points in $i$.
\par 
For each $j=1, ..., d$, we refer to $q_{0, j} \in (0, 1)$ in Equation (\ref{eq:bernoulli}), as the background rate of change-points in series $j$. We then let $W_{0, j} > 0$ be the weight assigned to the background change-point process. To indicate which pairs of series $(i,j)$ exhibit lead-lag relationships, we define an adjacency matrix with entries $A_{i, j} \in \{0, 1\}$. Then, for each edge $(i, j)$ we define an edge weight $W_{i, j} > 0$ and an edge impulse function $g_{i, j} : \mathbb{Z} \rightarrow [0, 1]$. Our model for $p_{j, t}(\bxp)$ is then given by 
\begin{equation}\label{eq:pjt}
    p_{j, t}(\bx_{t-1}) = Z_{j}^{-1} \left[W_{0, j}q_{0, j} + \sum_{i=1}^d A_{i, j}W_{i, j}g_{i, j}(x_{i, t-1})\boldsymbol{1}(x_{i, t-1} < t-1)\right]
\end{equation}
where $Z_{j}$ is a normalizing constant given by $Z_j = W_{0, j} + \sum_{i=1}^{d}A_{i, j}W_{i, j}$ and $\boldsymbol{1}(x_{i, t-1} < t-1)$ is the indicator function necessary for preventing the model treating $t=0$ as an observed change-point. We can also write the model in terms of a set of normalized weights as
\begin{equation} \label{eq:pjt_norm}
    p_{j, t}(\bxp) = W_{0, j}^*q_{0, j}^* + \sum_{i=1}^d W_{i, j}^*g(x_{i, t-1})\boldsymbol{1}(x_{i, t-1} < t-1)
\end{equation}
with $W_{0, j}^* = W_{0, j}/Z_j$ and $W_{i, j}^* = A_{i, j}W_{i, j} / Z_j$ for $i=1, ..., d$. Since we are interested in uncovering the latent network structure of the multivariate series we use the model representation in Equation \eqref{eq:pjt}. Doing so allows us to directly study the effect the edge adjacency matrix has on the lead-lag dynamics between the series. 
\par 
Equation \eqref{eq:pjt} presents the probability of series $j$ transitioning to a change-point, $p_{j, t}(\bx_{t-1})$, as a weighted sum of the background rate $q_{0, j}$ and a linear combination positive \emph{impulses} which depend on the distance to the most recent change-point in series $i$ such that $A_{i, j} = 1$. The magnitude of each impulse is controlled by the edge weight $W_{i, j}$ and its shape controlled by the edge impulse function $g_{i, j}(x_{i, t-1})$. However, if $A_{i, j} = 0$ for all $i = 1, ..., d$ then $p_{j, t}(\bxp)$ defaults back to the independent Bernoulli process of Equation \eqref{eq:bernoulli}. 
\par 

Due to the term $\boldsymbol{1}(x_{i, t-1} < t-1)$, introduced to prevent $t=0$ from being treated as a change point, our Markov process has inhomogeneous transitions. However, after all series have hit their first change-points, the chain is governed by the following homogeneous transitions
\begin{equation} \label{eq:pjt_homo}
    p_{j}(\bxp) = Z_j^{-1} \left[W_{0, j}q_{0, j} + \sum_{i=1}^dA_{i, j}W_{i, j}g_{i, j}(x_{i, t-1})\right]
\end{equation} 
The choice of edge impulse function, $g_{i, j}(x)$, is an important one. This function is analogous to those seen in the network Hawkes process literature \parencite[]{Linderman2014}. Functions with peaks close to one encourage change-points to occur shortly after those in leading series. Whereas functions with ``humps'' away from one can be used to model more distant dependence. In this article we restrict our attention to the case where we expect short delays between change-points. We therefore specify $g_{i,j}(t)$ to be the PMF of the geometric distribution
\begin{equation}\label{eq:geometric}
    g_{i, j}(t) = q_{i, j}(1-q_{i, j})^{t-1}
\end{equation}
where parameter $q_{i, j} \in (0, 1)$ controls the rate of decay. Larger values of $q_{i, j}$ will result in a larger mass near one, promoting short delays, while values nearer to zero will flatten the function, encouraging longer delays. 
\par
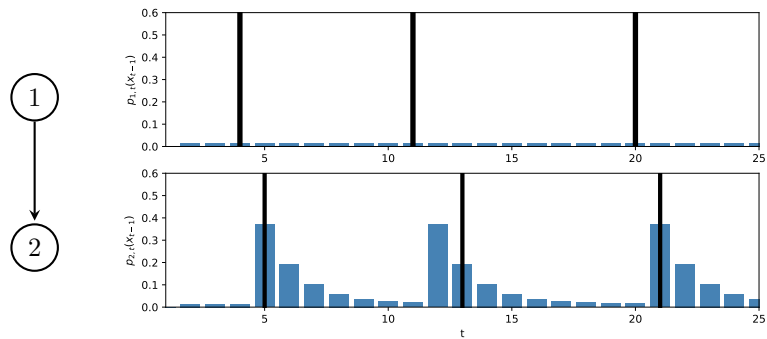
\begin{figure}[H]
  \centering
    \begin{minipage}[t]{0.7\textwidth}
    
    \begin{minipage}[t]{0.2\textwidth}
      \centering
      \begin{tikzpicture}[>=stealth,shorten >=1pt,auto,
                        thick,main node/.style={circle,draw}]
          \node[main node] at (0, 2)   (1) {1};
          \node[main node] at (0, 0)   (2) {2};
          \node[main node, opacity=0.0] at (0, -1)   (3) {2};
          \draw[->] (1) to (2);
        \end{tikzpicture}
    \end{minipage}%
    \hfill
    \begin{minipage}[t]{0.8\textwidth}
      \centering
      \includegraphics[width=\textwidth]{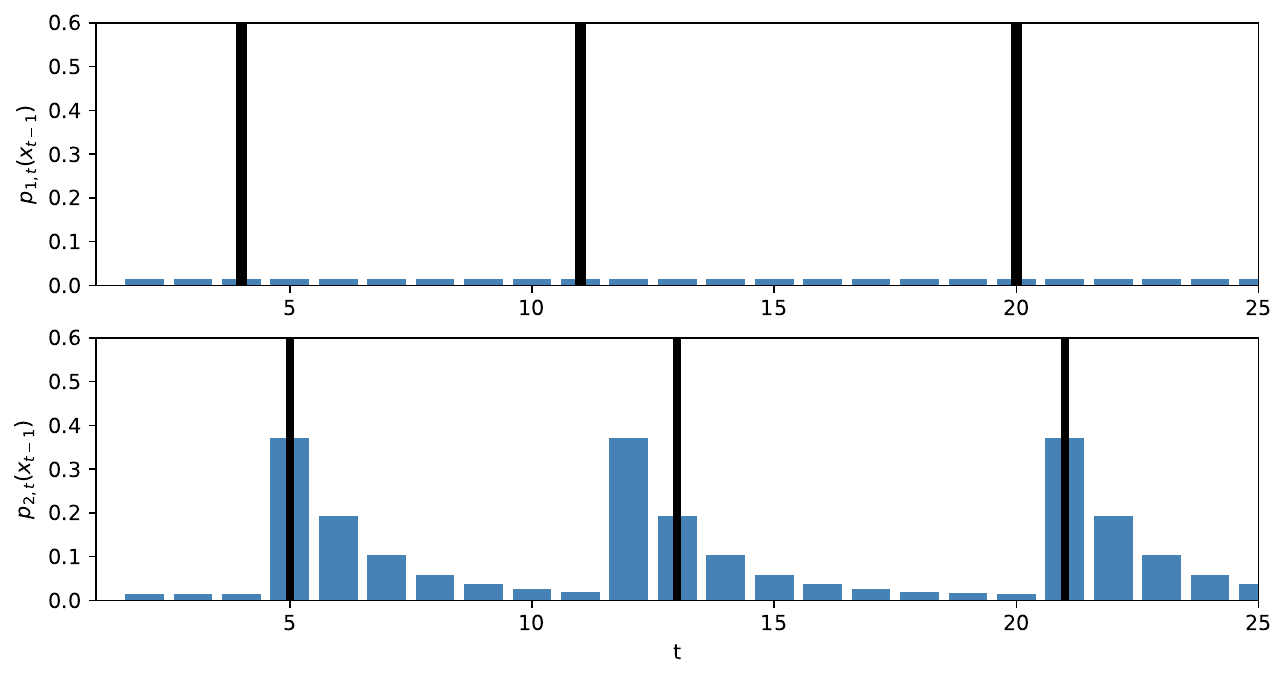}
    \end{minipage}
    \end{minipage}
  \caption{Simulation from the NetCP process (right) for $d=2$ with $\bA$ representing a chain graph (left) with $A_{1, 2} = 1$. Blue bars show the change-point probabilities, $p_{j, t}(\bxp)$, and solid lines the change-point locations (where $X_{j, t} = 1$). We use the parameters $q_{0 ,1}=q_{0,2} = 0.1$, $W_{0, 1} = W_{0, 2} = 1$, $W_{1, 2} = W_{2, 1} = 5$ and $q_{1, 2} = q_{2, 1} = 0.5$.}
  \label{fig:sim_dag}
\end{figure}

 
The construction of $p_{j, t}(\bxp)$ induces a form of mutual excitation between the change-point processes for each series. This is illustrated in Figure \ref{fig:sim_dag} which shows a realization of the prior for $d=2$ and $\bA$ representing a chain graph with single edge $(1, 2)$. Here, a change-point in series one causes an impulse in the change-point probabilities in series two. This impulse decays geometrically according to $g_{1, 2}(\cdot)$ and encourages change-points to occur in series two shortly after.

\subsection{Properties of the Prior}\label{sec:properties}
In this section we examine key properties of our prior, including stationarity, the expected number of change-points and the implied cross correlation between the change-point locations in different series. A clear understanding of the properties of our prior model can guide our specification of hyperpriors for parameters $(\bW_0, \bW, \bA, \boldsymbol{q}_0, \boldsymbol{q})$ that affect the lead-lag dependence of change points between series.
\par
As noted in Section \ref{sec:cp_model}, the hidden state vector, $\bX_t = (X_{1, t}, ..., X_{d, t})^{'}$, is initially governed by inhomogeneous transitions given by Equation \eqref{eq:pjt}. These transitions imply $$W_{0, j}q_{0, j}/Z_j \leq p_{j, t}(\bxp) < 1,$$  and so the rate at which the change points occur in series $j$ is greater than or equal to that of a Bernoulli process with parameter $W_{0, j}q_{0, j} / Z_j$. A trivial upper bound on the expected arrival time of the first change-point in series $j$ can then be obtained as
\begin{equation}
    \mathbb{E}(\tau_{j, 1}) \leq Z_j /W_{0, j}q_{0, j}
\end{equation}
Once all series experience their first change-point, the transitions become homogeneous, see Equation \eqref{eq:pjt_homo}, and so investigating the limiting behavior of NetCP is equivalent to studying the behavior of the homogeneous phase. 
\begin{proposition}\label{prop:one}
Let $\bX_t$ be a Markov process defined by the homogeneous transitions given by Equation \eqref{eq:pjt_homo}. Then, for any initial condition, $\bX_t$ converges to a unique stationary distribution.
\par
\noindent
\end{proposition}
\noindent
Proposition \ref{prop:one} tells us that the homogeneous phase of the process is stationary and therefore does not exhibit degenerate or divergent behavior. Therefore, the limiting distribution of the NetCP process is stable and is equal to the stationary distribution of the homogeneous phase. The asymptotic stationarity of the chain guarantees that the run lengths, $\bX_t = (X_{1, t}, ..., X_{d, t})^{'},$ do not degenerate at 1 (constant change-points) or diverge to infinity (never having change-points). 
\par 
However, the choice of graph structure and graph parameters $(\bW_0, \bW, \bA, \boldsymbol{q}_0, \boldsymbol{q})$ can cause the chain to produce many change-points in a given time frame. To study the properties of the change-point locations we introduce a set of binary random variables $$U_{j, t} = \boldsymbol{1}(X_{j, t} = 1) \in \{0, 1\}$$ indicating the locations where change-points occur in each series. We proceed by studying the behavior of the NetCP process when the structure of the graph is acyclic and cyclic.

\subsubsection{Acyclic Graphs}
In this subsection, we study the behavior of the NetCP when the underlying graph is acyclic. In order to build intuition for how the process behaves in this scenario, we consider the simple case of a chain graph with $d=2$ and a single edge $(1, 2)$. In this scenario, we have $A_{1, 1} = A_{2, 1} = 0$, and so the change-point process in series one evolves according to a Bernoulli process, independently of series two, with $Pr(U_{1, t} = 1) = q_{0, 1}$ for all $t > 1$. However, the presence of the edge $(1, 2)$ means that the rate of change-points in series two is dependent on series one. 

\begin{proposition}\label{prop:three}
Suppose $\bX_t$ is the two-dimensional process governed by the NetCP transitions given in \eqref{eq:pjt}. Suppose that $\bA$ represents a chain graph with $A_{1, 2} = 1$ and an edge impulse function, $g_{1, 2}(.),$ given by Equation \eqref{eq:geometric}. Then for $t > 1$
\begin{align}
    Pr(U_{2, t} = 1) = \frac{W_{0, 2}q_{0, 2} + W_{1, 2}\lambda(t-1, q_{0, 1}, q_{1, 2})}{W_{0, 2} + W_{1, 2}}
\end{align}
where
\begin{equation}
    \lambda(t, q_{0, 1}, q_{1, 2}) = q_{0, 1}q_{1, 2}\left[\frac{1 - [(1-q_{0,1})(1-q_{1, 2})]^{t-1}}{1 - (1-q_{0, 1})(1-q_{1, 2})}\right]
\end{equation} 
\end{proposition}
\noindent
From Proposition \ref{prop:three}, we can see that the marginal probability of a change-point in series two at time $t$ is a weighted combination of the background rate for series two and $\lambda(t-1, q_{0, 1}, q_{1, 2})$, which quantifies the contribution of the edge impulse function acting on the most recent change-point in series one. Note that $\lim_{q_{1, 2} \rightarrow1}\lambda(t, q_{0, 1}, q_{1, 2}) = q_{0, 1}$, suggesting that as we make the edge impulse function more peaked, $Pr(U_{2, t} = 1)$ becomes a simple weighted average of the background rates for the two series. For a sample of size $T$, the expected number of change-points in series two is then:
\begin{equation}
    \sum_{t=2}^T Pr(U_{2, t} = 1) = \frac{(T-1)W_{0, 2}q_{0, 2} + W_{1, 2}\sum_{t=2}^{T}\lambda(t-1, q_{0, 1}, q_{1, 2})}{W_{0, 2} + W_{1, 2}}
\end{equation}
which remains a linear combination of the background rate for series two and the contribution from the edge impulse function from series one. The limiting (stationary) probability of a change-point in series two is then given by 
\begin{equation}
    \lim_{t \rightarrow \infty} Pr(U_{2, t} = 1) = \frac{W_{0, 2}q_{0, 2} + W_{1, 2}\lambda(q_{0, 1}, q_{1, 2})}{W_{0, 2} + W_{1, 2}}
\end{equation}
where $\lambda(q_{0, 1},q_{1, 2}) = q_{0, 1}q_{1, 2} / [1 - (1-q_{0, 1})(1-q_{1, 2})]$. \smallskip
\par 
The key motivation for the construction of the NetCP process was to induce lead-lag dependence across the change-point locations in different series. To quantify this dependence we compute the cross-autocovariance and cross-autocorrelation in the change-point locations. This is formalized in the next proposition for the chain graph in $d=2$.

\begin{proposition}\label{prop:four}
Suppose $\bX_t$ is the two-dimensional process governed by the NetCP transitions given in Equation \eqref{eq:pjt}. Suppose that $\bA$ represents a chain graph with $A_{1, 2} = 1$ and edge impulse function, $g_{1, 2}(.),$ given by Equation \eqref{eq:geometric}. Then for $t > 1$ and $h \geq 1$
\begin{equation} \label{eq:cov}
    \text{Cov}(U_{1, t}, U_{2, t+h}) = \frac{q_{0,1}W_{1, 2}[\lambda^*(h, q_{0, 1}, q_{1, 2}) - \lambda(t+h-1, q_{0, 1}, q_{1, 2})]}{W_{0, 2} + W_{1, 2}}
\end{equation} 
where $\lambda^*(t, q_{0, 1}, q_{1, 2}) = \lambda(t, q_{0, 1}, q_{1, 2}) + q_{12}[(1-q_{0, 1})(1-q_{1, 2})]^{t-1}$ and $\lambda(t, q_{0, 1}, q_{1, 2})$ given in Proposition \ref{prop:three}. 
\end{proposition}
Combining Proposition \ref{prop:four} with $Pr(U_{2, t} = 1)$ from Proposition \ref{prop:three} and $P(U_{1, t}=1)=q_{0, 1}$ we obtain the cross-autocorrelation. The stationary cross-autocorrelation can then be calculated by replacing $\lambda(t+h-1, q_{0, 1}, q_{1, 2})$ with its limit defined by $\lambda(q_{0, 1}, q_{1, 2})$.

\begin{figure}[H]
    \centering
    \includegraphics[width=\linewidth]{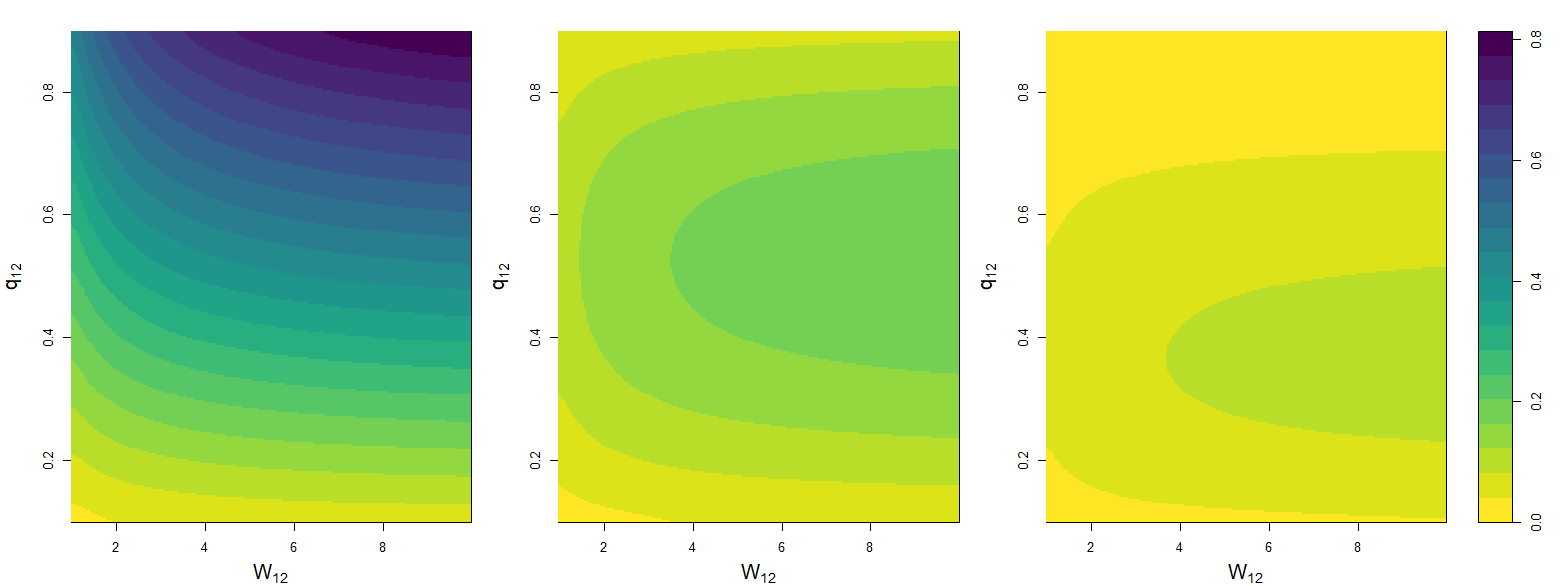}
    \caption{Stationary cross-autocorrelation of change-point indicators $\text{Cor}(U_{1, t}, U_{2, t+h})$ under the NetCP process for $d=2$ with $A_{1, 2} = 1$ for $h=1$ (left), $h=2$ (middle) and $h=3$ (right). Each plot displays a heat-map of the correlation induced by varying edge weight $W_{1, 2}$ and geometric edge impulse function parameter $q_{1, 2}$. We set the parameters $q_{0, 1} = q_{0, 2} = 0.1$ and $W_{0, 2} = 1$.}
    \label{fig:corr_heatmaps}
\end{figure}

Figure \ref{fig:corr_heatmaps} displays the stationary cross-autocorrelation, $\text{Cor}(U_{1, t}, U_{2, t+h})$, of the NetCP process using a chain graph in $d=2$, for varying $h$ and $(q_{1, 2}, W_{1, 2})$. Here, we can see that the cross-autocorrelation is highest at lag $h=1$ and decreases as the lag increases. At $h=1$, the correlation increases both in $q_{1, 2}$ and in $W_{1, 2}$. However, at lags $h=2$ and $h=3$, the highest correlation is achieved by setting lower values of $q_{1, 2}$. Recall that $q_{1,2}$ controls the flatness of the geometric edge impulse function $g_{1, 2}(x_{1, t-1})$. Smaller values of $q_{1, 2}$ result in a flatter edge impulse function, which places more mass on change-points occurring at longer lags. Therefore when the edge impulse function is the PMF or a geometric distribution, smaller values of $q_{i, j}$ should be used to obtain higher correlation at longer lags between series $i$ and $j$.

\subsubsection{Cyclic Graphs}
\begin{figure}[H]
  \centering
    \begin{minipage}[t]{0.7\textwidth}
    
    \begin{minipage}[t]{0.2\textwidth}
      \centering
      \begin{tikzpicture}[>=stealth,shorten >=1pt,auto,
                        thick,main node/.style={circle,draw}]
          \node[main node] at (0, 2)   (1) {1};
          \node[main node] at (0, 0)   (2) {2};
          \node[main node, opacity=0.0] at (0, -1)   (3) {2};
          \draw[->] (1) to[bend left] (2);
          \draw[->] (2) to[bend left] (1);
        \end{tikzpicture}
    \end{minipage}%
    \hfill
    \begin{minipage}[t]{0.8\textwidth}
      \centering
      \includegraphics[width=\textwidth]{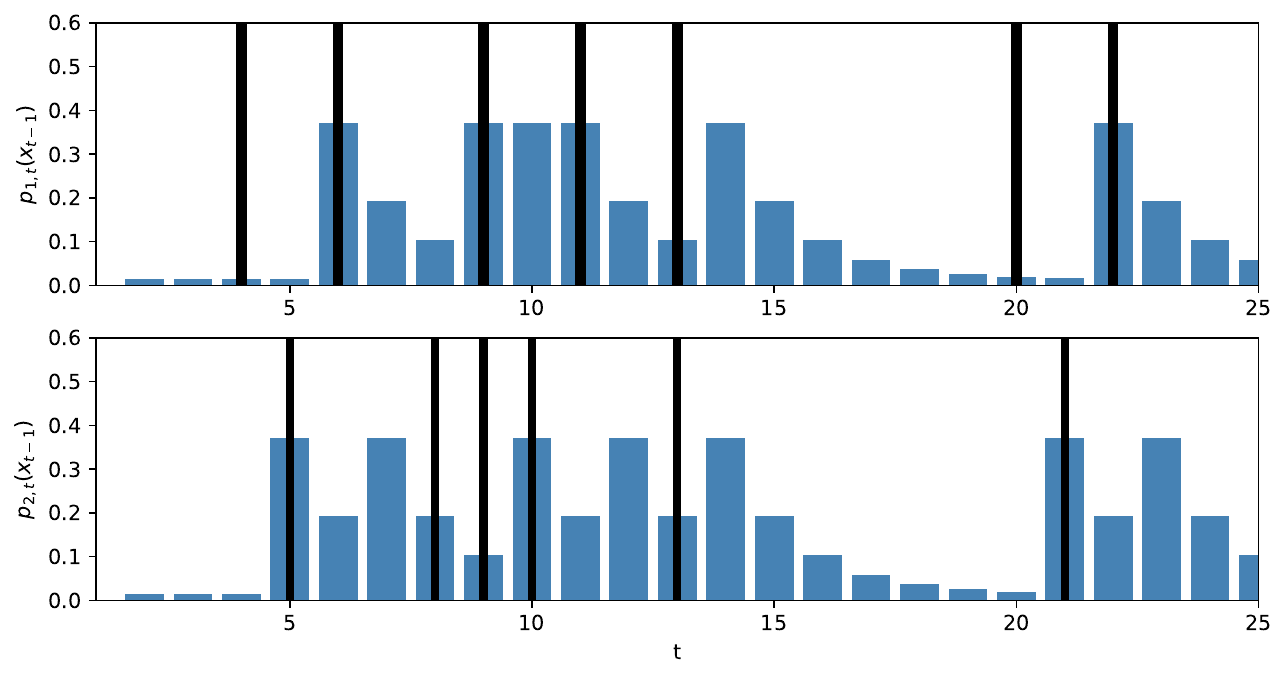}
    \end{minipage}
    \end{minipage}
  \caption{Simulation of the NetCP process (right) for $d=2$ with $\bA$ representing a cyclic graph (left) with $A_{1, 2} = A_{2, 1} = 1$. Blue bars show the change-point probabilities, $p_{j, t}(\bxp)$, and solid lines the change-point locations (where $X_{j, t} = 1$). We use the parameters $q_{0 ,1}=q_{0,2} = 0.1$, $W_{0, 1} = W_{0, 2} = 1$, $W_{1, 2} = W_{2, 1} = 5$ and $q_{1, 2} = q_{2, 1} = 0.5$.}
  \label{fig:sim_cycle}
\end{figure}
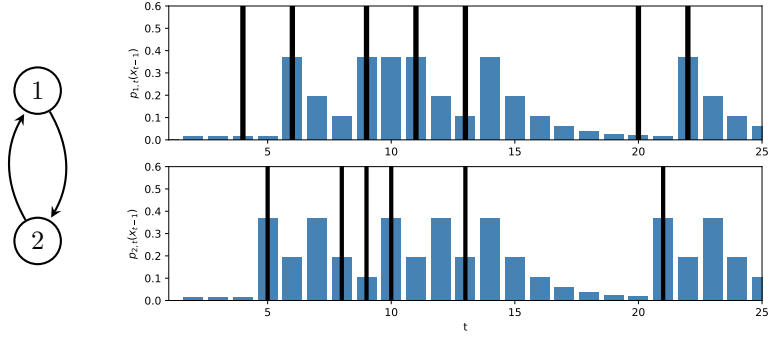

The mutual excitation observed for acyclic graphs induces cross-autocorrelation between pairs of series, resulting in the lead-lag behavior we wish to model. Cyclic graphs, however, result in positive feedback loops, whereby a change-point in a given series $j$ can trigger a change in a dependent series $i$ which can in turn trigger another change in $j$. This effect is illustrated for $d=2$ in Figure \ref{fig:sim_cycle} where the graph contains the cycle $\{(1, 2), (2, 1)\}$. Here, bursts of change-points become likely, whereby a change-point in series one causes an impulse in series two, which then feeds back into series one and vice versa. 

It is not hard to verify that if $\bA_{\text{D}}$ is the adjacency matrix of the acyclic graph with single edge $(1, 2)$ and $\bA_{\text{C}}$ the cyclic graph with edges $\{(1, 2), (2, 1)\}$, then $\text{Cov}(U_{1, t}, U_{1, t+h}|\bA_{\text{D}}) = 0$ and $\text{Cov}(U_{1, t}, U_{1, t+h}|\bA_{\text{C}}) > 0$. So under the cyclic graph, the indicators within a given series are positively correlated, suggesting bursts of consecutive change-points are likely. Bursts of change-points, as seen in Figure \ref{fig:sim_cycle}, are generally undesirable since in many applications a structural change tends to persist for some time. A prior which encourages many change-points in quick succession may therefore be prone to overfitting. For this reason, we recommend avoiding cyclic graph structures as discussed in Section \ref{sec:hyperpriors}.

\subsection{Hyperpriors} \label{sec:hyperpriors}
Motivated by our analysis of the NetCP process in Section \ref{sec:properties}, we now specify priors over the processes parameters $(\bW_0, \bW, \bA, \boldsymbol{q}_0, \boldsymbol{q})$. A standard prior used to model random graphs is the Erd\H{o}s-R\'enyi (ER) model \parencite[]{ER}, which specifies a Bernoulli prior for the adjacency matrix elements, $A_{i, j} \overset{\mathrm{iid}}{\sim}Bern(\rho)$ for $\rho \in (0, 1)$. This prior assigns equal probability, $\rho$, to each edge being present in the graph. However, our analysis in Section \ref{sec:properties} suggests that cyclic graphs can result in undesirable behavior in the change-point locations (bursts of change-points) and so we want our prior to assign little or no mass on such graphs.
\par
Constructing a prior which assigns non-zero mass only to directed acyclic graphs is difficult due to the space being combinatorially complex and highly constrained by acyclicity. We instead base our prior on a modification of the ER model which does not allow self edges or parallel edges. Specifically, we let $\rho \in (0, 1)$, and define our prior on $\bA$ as 
\begin{equation} \label{eq:prior_A}
    Pr(\bA=\boldsymbol{a}|\rho) = \prod_{j=1}^dPr(A_{j, j} = a_{j, j})\prod_{j=1}^d\prod_{i=j+1}^dPr(A_{i, j}=a_{i, j}, A_{j, i}=a_{j, i}|\rho)
\end{equation}
where $Pr(A_{j, j} = 0) = 1$ and 
\begin{equation} \label{eq:prior_A_pairs}
    Pr(A_{i, j}=a_{i, j}, A_{j, i}=a_{j, i}|\rho) = \begin{cases}
        \rho/2,\;\;\;\;\;\;\; \text{  if } a_{i, j} = 0, a_{j, i} = 1\\
        \rho/2,\;\;\;\;\;\;\; \text{  if } a_{i, j} = 1, a_{j, i} = 0\\
        1-\rho,\;\;\;\;\; \text{  if } a_{i, j} = 0, a_{j, i} = 0\\
        0,\;\;\;\;\;\;\;\;\;\;\; \text{  if } a_{i, j} = 1, a_{j, i} = 1\\
    \end{cases}
\end{equation}
In other words, for each pair of distinct vertices $(i, j)$ with $i \ne j$, a single directed edge is included with probability $\rho$, and excluded with probability $1 - \rho$. Under this choice, the expected number of edges is $\rho d(d-1)/2$, allowing us to control the sparsity of the graph through our prior on $\rho$. To encourage sparse graphs and prevent overfitting, we place a uniform prior with parameters $(0, 0.2)$ on $\rho$.
\par 
We choose independent Gamma priors for the background and edge weights, $(\boldsymbol{W}_0,\bW)$, i.e.
\begin{eqnarray}
    W_{0, j} &\sim& Ga(1, 1) \;\;\; \text{for } j=1, ...,d \\
    W_{i, j} &\sim& Ga(1, 1) \;\;\; \text{for } i=1, ...,d; \;j=1, ..., d
\end{eqnarray}
where $Ga(a, b)$ denotes the Gamma distribution with shape $a$ and rate $b$. We find that this choice works well when modeling the dependence between pairs of lead-lag series. Under this choice, we are centering the weights around one, producing correlations in the lagged change-point locations less than 0.5, see Figure \ref{fig:corr_heatmaps}. This choice has the added property that the induced prior on the vector of normalized weights for series $j$ is 
\begin{equation*}
    (W^*_{0, j}, W_{1, j}^*, ..., W_{d, j}^*) | \bA \sim Dir(1, A_{1, j}, ..., A_{d, j})
\end{equation*}
where $Dir(\alpha_1, ..., \alpha_K)$ is the $K$ dimensional Dirichlet distribution, with $\alpha_k=0$ indicating that the $k^{th}$ element in the sampled vector is zero almost surely. 
\par
Recall that the parameters $q_{i, j} \in (0, 1)$ control how quickly the correlation between a change-point in series $i$ and a subsequent change-point in series $j$ decays over time. Larger values result in a faster decay rate, indicating a change-point in the $j$th series is most likely to immediately follow a change-point in the $i$th series. Smaller values of $q_{i, j}$ allow for a longer delay. We choose the standard uniform prior:
\begin{equation}
    q_{i, j} \sim \text{Uniform}(0, 1) \;\;\;\text{for } i=1, ...,d; \;j=1, ..., d
\end{equation}
to reflect the absence of any prior information on the decay rate. For this reason we also choose the standard uniform prior for the background rates $q_{j, 0}$ which control the overall rate at which change-points occur in each series.

\section{Posterior Inference} \label{sec:post}
We now describe how to perform posterior inference for the NetCP model introduced in Section \ref{sec:model}. Our MCMC sampler extends the work of \textcite{Fearnhead2006} and \textcite{Whiteley2011} to the multivariate setting. We show how the hidden states for each series can be efficiently sampled in blocks using the forward-backward recursions of \textcite{Fearnhead2006}. For long time series, we demonstrate how particle MCMC methods \parencite[]{Andrieu2010, Whiteley2011} can be used to control computational cost while preserving a high level of sampling efficiency. The full joint posterior density is given by 
\begin{align}
    \pi(\bx, \boldsymbol{W}_0, \boldsymbol{W}, \boldsymbol{A}, \boldsymbol{q}_0, \boldsymbol{q}, \rho|\boldsymbol{y}) &\propto f(\boldsymbol{y}|\boldsymbol{x}) Pr(\bX=\bx|\boldsymbol{W}_0, \boldsymbol{W}, \boldsymbol{A}, \boldsymbol{q}_0, \boldsymbol{q}) \\ &\times f(\boldsymbol{W}_0)f(\boldsymbol{W})f(\bA|\rho)f(\rho)f(\boldsymbol{q}_0)f(\boldsymbol{q}) \nonumber.
\end{align}
The main difficulty arises when sampling the hidden states, $\bX$. One approach is to sample each $X_{j, t}$ one at a time from its full posterior conditional \parencite[see][]{Barry1993, Quinlan2024}. Since each $X_{j, t}$ is either $1$ (a change-point) or $X_{j, t-1} + 1$ (no change-point), this can be performed with simple Bernoulli samples. While easy to implement, strong dependencies between the $X_{j, t}$ mean that this method tends to mix poorly \parencite[see][]{Fan2017}. Another popular approach is to use reversible jump MCMC \parencite[]{Green1995}, but again this is sub-optimal since designing efficient proposal moves is non trivial.
\par
We construct an efficient sampling scheme by sampling the hidden states in each series as a block. This may be done using the method of \textcite{Fearnhead2006}, which employs a set of forward-backward recursions. However, these recursions have a computational cost of $\mathcal{O}(T^2)$ per series, making them impractical within an MCMC framework. To balance sampling efficiency with computational cost, we instead use the particle approximation of \textcite{Whiteley2011}, maintaining the correct stationary distribution via conditional particle filtering within a particle Gibbs framework \parencite[]{Andrieu2010}. A single iteration of our sampling scheme involves the following steps:
\begin{enumerate}
    \item For $j=1, ..., d$, sample $\bx_{j, 0:T}$ from $\pi(\bx_{j, 0:T}|\boldsymbol{y}, \bx_{(-j), 0:T}, \cdots)$ via a particle Gibbs step.
    \item Sample $(\boldsymbol{W}_0, \boldsymbol{W}, \boldsymbol{A}, \boldsymbol{q}_0, \boldsymbol{q}, \rho)$ from their full posterior conditionals.
\end{enumerate}
We sample the static parameters in step 2 using standard Gibbs/Metropolis steps, the details of which can be found in Appendix B.
\par
For the remainder of this section we discuss our blocked sampling scheme for the hidden states $\bx_{j, :0:T}$ for $j=1, ..., d$. Step 1 requires us to sample from the joint PMF of the hidden states in a given series $j$ that is,
\begin{eqnarray} \label{eq:block_joint}
    \pi(\bx_{j, 0:T}|\boldsymbol{y}, \bx_{(-j), 0:T}, \cdots) &\propto& Pr(\bX_{j, 0:T} = \bx_{j, 0:T}|\bX_{(-j), 0:T} = \bx_{(-j), 0:T}) f_j(\boldsymbol{y}_{j, 0:T}|\bx_{j, 0:T}).
\end{eqnarray} 
\noindent
Observe that due to the Markov properties of $\bX_{t}$, the term $Pr(\bX_{j, 0:T} = \bx_{j, 0:T}|\bX_{(-j), 0:T} = \bx_{(-j), 0:T})$ can be factorized as
\begin{equation}\label{eq:full_cond}
    Pr(\bX_{j, 0:T} = \bx_{j, 0:T}|\bX_{(-j), 0:T} = \bx_{(-j), 0:T}) \propto Pr(X_{j, 1} = x_{j, 1}) \prod_{t=2}^T P_{t}^{(j)}(x_{j, t}|x_{j, t-1}, \bx_{(-j), 0:T})
\end{equation}
where the term $P_{t}^{(j)}(x_{j, t}|x_{j, t-1}, \bx_{(-j), 0:T})$ defines a transition PMF for $(X_{j, t}|X_{j, t-1}, \bX_{(-j), 0:T})$ and is defined up to proportionality as
\begin{align}
    \label{eq:cond_trans}
    &P_{t}^{(j)}(x_{j, t}|x_{j, t-1}, \bx_{(-j), 0:T}) \propto \\ \nonumber
    &\begin{cases}
        Pr(X_{j, t} = x_{j, t}|\bX_{t-1} = \boldsymbol{x}_{t-1}^{[j=x_{j, t-1}]}) \prod_{\substack{i=1 \\ i \neq j}}^d Pr(X_{i, t+1} = x_{i, t+1}|\bX_{t} = \boldsymbol{x}_{t}^{[j=x_{j, t}]}), \;\;\; \text{for } 2 \leq t < T \\
        \Pr(X_{j, t} = x_{j, t}|\bX_{t-1} = \boldsymbol{x}_{t-1}^{[j=x_{j, t-1}]}), \;\;\;\;\;\;\;\;\;\;\;\;\;\;\;\;\;\;\;\;\;\;\;\;\;\;\;\;\;\;\;\;\;\;\;\;\;\;\;\;\;\;\;\;\;\;\;\;\;\;\;\;\;\;\;\;\;\;\;\;\;\;\;\;\;\;\; \text{for } t = T
    \end{cases}
\end{align}
where  $\bx^{[j=s]} = (x_1, ..., x_{j-1}, s, x_{j+1}, ..., x_{d})^{T}$, denotes the vector $\bx$ with the value $s$ imputed at the $j^{th}$ index. The hidden state at time $t$ can take two values ($1$ or $x_{j, t-1}+1$), and so the normalizing constant of $P_{t}^{(j)}(x_{j, t}|x_{j, t-1}, \bx_{(-j), 0:T})$ can be easily found. Therefore, conditional on the hidden states in all other series, the hidden states in series $j$ follow Markov a chain, with transitions given by Equation (\ref{eq:cond_trans}).
Recall that the likelihood for series $j$ can be factorized as
\begin{eqnarray} \label{eq:like_factorised}
    f_j(\boldsymbol{y}_{j, 0:T}|\bx_{j, 0:T}) = \prod_{t=1}^{T}f_j(y_{j, t}|x_{j, t}, \boldsymbol{y}_{j, 0:t-1})
\end{eqnarray}
where $f_j(y_{j, t}|x_{j, t}, \boldsymbol{y}_{j, 0:t-1})$ is given by Equation (\ref{eq:pred_like}). The full posterior conditional, Equation (\ref{eq:block_joint}), can therefore be written as
\begin{equation} \label{eq:block_joint _factorised}
    \pi(\bx_{j, 0:T}| \boldsymbol{y}, \bx_{(-j), 0:T}, \cdots) \propto f_j(y_{j, 1}) \prod_{t=2}^T P_t^{(j)}(x_{j, t}|x_{j, t-1}, \bx_{(-j), 0:T}) f_j(y_{j, t}|x_{j, t}, \boldsymbol{y}_{j, 0:t-1})
\end{equation}
where $x_{j, 1} = 1$ as per Equation (\ref{eq:initial_state}). 
\par
The manner in which Equation (\ref{eq:block_joint _factorised}) factorizes allows us to use the forward-backward recursions of \textcite{Fearnhead2006} to perform direct sampling. However, due to the $\mathcal{O}(T^2)$ computational time needed for the exact recursions, we instead use the method of \textcite{Whiteley2011} to replace the exact filtering distributions at time $t$ with an approximation. This approximation consists of $N$ weighted particles and reduces the computational cost to $\mathcal{O}(NT)$. A key component in the method of \textcite{Whiteley2011} is the particle filter of \textcite{Fearnhead2007}. It differs from standard particle methods in two ways. First, since the state space is discrete, a deterministic search of the proposal space is used instead of a random proposal, a strategy which greatly reduces Monte Carlo error. Second, instead of using multinomial resampling, it employs Stratified Optimal Resampling (SOR), which is shown to be optimal over all existing unbiased resampling methods in terms of the maximum Kolmogorov–Smirnov distance. This approach also ensures that no duplicate particles are stored at each iteration, since there is no advantage in storing multiple copies in a discrete space \parencite[see][]{Fearnhead2007}. 
\par
Algorithm \ref{alg:particle_gibbs} describes the resulting particle Gibbs update for $\bx_{j, 0:T}$. In the following, we denote the set of particles at time $t$ by $\mathcal{S}_t \subseteq \{1, ..., t\}$ and their associated weights by $\mathcal{W}_t = \{w_t(s_t)\}_{s_t \in S_t}$. Therefore, at time $t$, the particle $s \in \mathcal{S}_t$ has weight $w_t(s)$.
\begin{algorithm}[H]
\caption{Particle Gibbs update of $\bx_{j, 0:T}$}
\label{alg:particle_gibbs}
\begin{algorithmic}[1]
\Require $\bx^*$ (current hidden state matrix), $N$ (number of particles).
\State At time $t=1$, set $\mathcal{S}_1 = \{1\}$ and $\mathcal{W}_1 = \{1\}$.
\For{$t=2, ..., T$}
    \State Set $\mathcal{S}_t = \{s_{t-1} + 1: s_{t-1} \in \mathcal{S}_{t-1}\} \cup \{1\}$.
    \For{each $s_{t-1} \in \mathcal{S}_{t-1}$}
        \State Set $w_t(s_{t-1}+1) \propto w_{t-1}(s_{t-1}) P_t^{(j)}(s_{t-1}+1|s_{t-1}, \bx_{(-j), 0:T}^*) f_j(y_{j, t}|s_{t-1}+1, \boldsymbol{y}_{j, 0:t-1})$
    \EndFor
    \State Set $w_t(1) = f_j(y_{j, t}) \sum\limits_{s_{t-1}\in \mathcal{S}_{t-1}} w_{t-1}(s_{t-1}) P_t^{(j)}(1|s_{t-1}, \bx_{(-j), 0:T}^*)$
    \State Normalize the weights $\mathcal{W}_t$ such that $\sum\limits_{s_t\in\mathcal{S}_t} w_t(s_t) = 1$.
    \State If $|\mathcal{S}_t| > N$, resample $(\mathcal{S}_t, \mathcal{W}_t)$ using the Conditional SOR algorithm (conditional on $x_{j, t}^*$).
\EndFor
\State Given $\{(\mathcal{S}_t, \mathcal{W}_t)\}_{t=1, ...,T}$ generate a new sample $\bx_{j, 0:T}$ using the backward sampling algorithm.
\end{algorithmic}
\end{algorithm}
\noindent
Algorithm \ref{alg:particle_gibbs} requires additional algorithms to perform the resampling in step 9 and backward sampling in step 11. These algorithms are the same as those given in \textcite{Whiteley2011} and can be found in Appendix B.
\par
Note that if the number of particles $N$ is set to $T$, then we never perform resampling in step 9 and Algorithm \ref{alg:particle_gibbs} is equivalent to the exact filtering recursions of \textcite{Fearnhead2006}. However, when $N < T$, the sampler maintains the correct stationary distribution due to conditioning on the previous sample \parencite[]{Andrieu2010}. In our simulated and real datasets we find that relatively few particles ($N \approx 200$) are required to obtain mixing similar to that of the direct sampling method where $N=T$. On a small simulated dataset, we find that the particle Gibbs with only $N=50$ obtains an integrated autocorrelation time that is 49 times smaller than that of the single-site Gibbs sampler which updates $X_{j, t}$ one at a time. Details of this analysis can be found in Appendix C. 
\par
A key assumption in Algorithm \ref{alg:particle_gibbs} is that the segment specific parameters may be marginalized out of the model, producing a Rao-Blackwellized filtering scheme. This results in highly efficient Monte Carlo estimates although restricts us to specifying conjugate segment likelihood models. The approach could be extended to handle NetCP models with non-conjugate segment densities by replacing the particle filter of \textcite{Fearnhead2007} with that of \textcite{Chopin2006}. Here, the segment specific parameters may be left in the model and included in the filtering steps, albeit at an increased computational cost and a reduction in Monte Carlo efficiency. 
\par 
Finally, we point out that it is trivial to modify the above particle Gibbs step to sample from the posterior of multivariate Markov chains other than the change-point model presented here. This is possible as long as the transition PMF is in the form given in Equation (\ref{eq:transition}) and the likelihood factorizes according to Equation (\ref{eq:joint_like}). The updating in steps 3-7 of Algorithm \ref{alg:particle_gibbs} can be replaced with the more general particle filter of \textcite{clifford2003}, allowing for Markov models which could re-enter previous states.

\section{Illustrations}\label{sec:illus}
In this section we evaluate the performance of NetCP in modeling change-points and inferring the underlying graph structure using both simulated and real data examples. 

\subsection{Simulation Study} \label{sec:sim_study}
We first perform a simulation study to test \begin{enumerate}[label=(\roman*)]
    \item NetCP's ability to infer the true underlying graph structure. 
    \item NetCP's fit, in terms of log model evidence, against several competing priors.
\end{enumerate} 
We consider five scenarios for the change-point generating process and two possible segment likelihood models, resulting in 10 distinct data generating scenarios. In all cases, we use dimension $d=4$.
Our change-point scenarios are designed to test situations in which change-points in different series (a) exhibit lead-lag relationships, (b) occur simultaneously across all, or subsets of the series and (c) are independent. Below we provide a description of each scenario \textbf{S1} - \textbf{S5}.

\begin{enumerate}[label=\textbf{S\arabic*}]
    \item Simulate from the NetCP process with parameters $\boldsymbol{W}_0 = \boldsymbol{1}_4$ and $\boldsymbol{q}_0 = \frac{1}{40}\boldsymbol{1}_4$. We use an adjacency matrix corresponding to a chain graph with $A_{1, 2} = A_{2, 3} = A_{3, 4} = 1$ and all other $A_{i, j} = 0$. The edge weights are set to $W_{1, 2} = W_{2, 3} = W_{3, 4} = 5$ and the edge decay parameters to $q_{1, 2} = q_{2, 3} = q_{3, 4} = 0.6$. The correlations induced by this choice can be seen in Figure \ref{fig:corr_heatmaps}.
    \item Simulate from the NetCP process with the same parameters as \textbf{S1} but instead setting $A_{2, 3} = 0$, resulting in two sub-graphs.
    \item Simulate the change-points simultaneously in all series. The change-point locations follow a Bernoulli process with parameter $1/40$. 
    \item Simulate the change-points simultaneously in series 1 \& 2 according to a Bernoulli process with parameter $1/40$. The change-points in series 3 \& 4 are also generated simultaneously with an independent and identical Bernoulli process. 
    \item Change-points in each series are generated according to an independent Bernoulli process with parameter $1/40$.
\end{enumerate}
\par 
For each change-point scenario (\textbf{S1} - \textbf{S5}), we generate data for two likelihood models. The first is the AR model presented in Equation (\ref{eq:ar_likelihood}). Under this scenario, the data in each segment of each series follows an AR(1) process, and the segment parameters $(\phi_1, \sigma^2)$ alternate between four possible states $\{(-0.8, 0.3^2), (0.8, 1^2), (-0.8, 2^2), (0.8, 3^2)\}$ for all series. 
\par 
Many existing change-point models focus on detecting changes in means \parencite[]{Truong2020} and our second likelihood scenario is designed for this purpose. Under this scenario, the observations in each series are independent Gaussian with fixed variance $\sigma_j^2$ and piecewise constant means which are sampled independently from a $\mathcal{N}(0, \gamma_j^2)$ distribution. This likelihood model is conjugate and the corresponding marginal segment density can be found in Appendix D. For this scenario's simulated data we set $\sigma_j^2 = 0.5$ and $\gamma_j^2 = 3$ for all series $j$. For each scenario and likelihood model, we generate 50 datasets, each with $d=4$ time series and $T=500$ time points. 
\par 
To evaluate NetCP’s ability to infer the true underlying graph, we assess the model’s output for \textbf{S1}, \textbf{S2}, and \textbf{S5} (\textbf{S5} corresponds to NetCP with an empty graph). Specifically, we examine whether NetCP correctly predicts edges $(i, j)$ in the graph by comparing the posterior probabilities $Pr(A_{i,j} = 1 \mid \boldsymbol{y})$ to the true adjacency matrix for each scenario. Classification performance is quantified using the Area Under the Receiver Operating Characteristic (ROC) Curve (AUC), which plots the true positive rate against the false positive rate for varying thresholds of $Pr(A_{i,j} = 1 \mid \boldsymbol{y})$. AUC values are averaged over the 50 simulated datasets for each change-point scenario and likelihood model. However, for scenario \textbf{S5}, the true adjacency matrix contains only zeros (no edges), making the AUC undefined. In this case, we report the true negative rate with a threshold of 0.5, averaged over the 50 datasets.
\par 
When fitting the NetCP model on the AR process likelihood we set the hyperparameters $\alpha_j=\beta_j=\delta_{j, 1} = 1$ for all series $j$. For the Gaussian likelihood model we set the hyperparameters to the true values $\sigma_j^2 = 0.5$ and $\gamma_j^2 = 3$ for all series $j$. We then estimate the NetCP model using the particle Gibbs scheme with 150 particles and 5000 iterations, discarding the first 500 as burn-in.
\begin{table}[H]
    \centering
    \begin{tabular}{ccc}
        \cline{2-3}
         & \textbf{S1} & \textbf{S2}  \\
        \hline
        Gaussian Mean & 0.99 (0.84, 1.00) & 0.99 (1.00, 1.00)   \\
        AR Process &  0.97 (0.71, 1.00) & 0.97 (0.71, 1.00)  \\
        \hline
    \end{tabular}
    \caption{Mean and 5\% quantiles of the Area Under the ROC Curve (AUC) computed over 50 datasets for scenarios \textbf{S1} and \textbf{S2}, under both AR process and Gaussian mean likelihoods. Higher AUC values indicate better edge prediction performance.} 
    \label{tab:auc}
\end{table}
Table \ref{tab:auc} displays the AUC for the edge prediction test on scenarios \textbf{S1} and \textbf{S2}. For scenario \textbf{S5}, where AUC is undefined, we threshold the inclusion probabilities $Pr(A_{i, j} = 1|\boldsymbol{y})$ at 0.5, obtaining an average true negative rate of 0.998 and 0.987 for the Gaussian mean and AR Process likelihoods respectively. From this we can see that on a sample of $T=500$ time points, NetCP is able to nearly perfectly recover the true $4\times4$ adjacency matrix under each scenario.
\par 
For test (ii) we compare NetCP to three competing change-point priors. The first is the special case of NetCP obtained by setting $\bA = \boldsymbol{0}$, which is equivalent to the Bernoulli prior of \textcite{Yao1984} being applied to each series independently. We refer to this model as the ``Yao'' model. We also consider the two priors proposed by \textcite{Quinlan2024}. The first is the Non-Global Correlated Change-Point PPM, which we refer to as ``NG-CCP''. This prior is constructed by replacing the transition probabilities in Equation (\ref{eq:transition_j}) with
\begin{eqnarray}
    p_{j, t}(\bx_{t-1}) &=& p^*_{j, t}, \;\;\; \text{ for }j = 1, ..., d \\
    (\text{logit}(p^*_{1, t}), ..., \text{logit}(p^*_{d, t})) &\sim& t_d(\nu, \mu, \Sigma)
\end{eqnarray}
where $t_d$ denotes the $d$-dimensional t distribution. We also consider the ``Global'' prior of \textcite{Quinlan2024}, which we refer to as ``G-CCP'', obtained by setting 
\begin{eqnarray}
    p_{j, t}(\bx_{t-1}) &=& p^*_{j}, \;\;\; \text{ for }j = 1, ..., d \\
    (\text{logit}(p^*_{1}), ..., \text{logit}(p^*_{d})) &\sim& t_d(\nu, \mu, \Sigma).
\end{eqnarray}
These priors encourage simultaneous change-points across series by allowing the change-point probabilities for a fixed time to be correlated through choice of $(\nu, \mu, \Sigma)$. \textcite{Quinlan2024} do not provide priors for these parameters, however, they do provide a method to compute sensible default values. Both the G-CCP and NG-CCP models can be estimated using the particle Gibbs scheme described in Section \ref{sec:post} and the latent transition probabilities $p^*_{j, t}$ and $p^*_j$ sampled using the Metropolis steps given in \textcite{Quinlan2024}.
\par
In order to compare model performance, we use the MCMC output to estimate the log Bayes factor, relative to our NetCP model, for each competing model as
\begin{equation*}
    log\text{BF}_M = logf_M(\boldsymbol{y}) - logf_{NetCP}(\boldsymbol{y})
\end{equation*}
where $logf_M(\boldsymbol{y})$ is the log model evidence for model $M$. Positive values of $log\text{BF}_M$ support model $M$ over NetCP. Values in excess of five suggest strong support. Similarly negative values suggest support for NetCP. We compute $logf_M(\boldsymbol{y})$ as follows, 
\begin{equation*}
    log f_M(\boldsymbol{y}) = \sum_{t=1}^{T} logf_M(\boldsymbol{y}_t|\boldsymbol{y}_{0:t-1}) 
\end{equation*}
The posterior predictive likelihoods $f_M(\boldsymbol{y}_t | \boldsymbol{y}_{0:t-1})$ can be approximated by running the Gibbs sampler using the data up to time $t-1$; at each iteration, the hidden state $x_t$ is sampled, and the predictive likelihood is estimated by averaging the likelihood of observation $\boldsymbol{y}_t$ over these sampled hidden states. To fit each model, we produce 5000 posterior samples using 150 particles and 500 burn-in. To achieve this we use the King's College London (KCL) CREATE computing cluster.
\begin{table}[H]
    \centering
    \resizebox{0.95\textwidth}{!}{%
    \begin{tabular}{cccccc}
        \cline{2-6}
         & \textbf{S1} & \textbf{S2} & \textbf{S3} & \textbf{S4} & \textbf{S5} \\
        \hline
        NetCP &\textbf{0} & \textbf{0} & 0 & 0 & 0 \\
        Yao & -29.0 (-52.9, -6.2) & -17.2 (-30.3, -3.5) & -38.5 (-52.8, -17.9) & 2.6 (-2.5, 7.6) & 6.2 (4.6, 7.6)  \\
        G-CCP & -26.9 (-53.0, -2.9) & -15.1 (-28.4, 0.8) & -37.1 (-53.3, -14.7) & \textbf{4.2 (-2.4, 11.5)} & \textbf{7.5 (4.7, 10.0)}  \\
        NG-CCP & -41.8 (-85.6, -3.92) & -34.1 (-61.4, -8.8) & \textbf{2.8 (-14.1, 21.2)} & 1.4 (-11.9, 15.0) & -15.8 (-33.3, -3.9) \\
        \hline
    \end{tabular}
    }
    \caption{Average model log Bayes Factors relative to NetCP, with 5\% quantiles computed over 50 simulated datasets for each change-point scenario under the Gaussian means likelihood. The highest average Bayes Factor in each scenario is shown in bold. Values above zero indicate models outperforming NetCP.}
    \label{tab:model_ev_norm}
\end{table}
\noindent
Table \ref{tab:model_ev_norm} displays the averaged log Bayes factors, relative to NetCP, for each model under the Gaussian means likelihood scenario. In \textbf{S1} and \textbf{S2}, which involve lead-lag relationships in change-points, NetCP clearly outperforms all other methods, with substantial negative log Bayes factors observed for competing models. In \textbf{S3}, where change-points occur simultaneously across series, NG-CCP achieves the highest score relative to NetCP, though the interval overlaps zero, suggesting comparable performance. In \textbf{S4}, G-CCP performs best, though differences between methods are again modest. For \textbf{S5}, which contains independent change-points across series, G-CCP yields the highest average Bayes factor, but Yao performs similarly, as indicated by their overlapping intervals. Overall, these results suggest that NetCP excels in structured, temporally misaligned settings, while G-CCP and Yao are more competitive in scenarios with simultaneous or independent change-points.

\begin{table}[H]
    \centering
    \resizebox{0.95\textwidth}{!}{%
    \begin{tabular}{cccccc}
        \cline{2-6}
         & \textbf{S1} & \textbf{S2} & \textbf{S3} & \textbf{S4} & \textbf{S5} \\
        \hline
        NetCP & \textbf{0} & \textbf{0} & \textbf{0} & \textbf{0} & 0 \\
        Yao & -17.8 (-32.3, --1.4) & -9.1 (-22.6, 0.8) & -55.3 (-74.2, -39.3) & -21.0 (-27.9, -14.7) & 7.3 (5.4, 9.5)  \\
        G-CCP & -16.2 (-32.0, 1.2) & -8.2 (-21.1, 1.8) & -54.6 (-75.2, -36.8) & -20.1 (-27.4, -12.1) & \textbf{8.1 (5.3, 11.0)}  \\
        NG-CCP & -20.8 (-45.1, 2.3) & -21.7 (-39.3, -5.6) & -19.0 (-34.9, -2.8) & -25.2 (-41.6, -7.6) & -12.8 (-29.7, 1.4) \\
        \hline
    \end{tabular}
    }
    \caption{Average model log Bayes Factors relative to NetCP, with 5\% quantiles computed over 50 simulated datasets for each change-point scenario under the AR Process likelihood. The highest average Bayes Factor in each scenario is shown in bold. Values above zero indicate models outperforming NetCP.}
    \label{tab:model_ev_ar}
\end{table}

Table \ref{tab:model_ev_ar} displays the same information as Table \ref{tab:model_ev_norm} but for the AR process likelihood scenario. Here NetCP performs strongly in \textbf{S1} - \textbf{S4}, achieving the highest estimated log Bayes factors relative to all other methods. In \textbf{S1} and \textbf{S2}, which involve lead-lag relationships between change-points, alternative methods show moderately negative Bayes factors, with some intervals overlapping zero, suggesting marginally weaker but not clearly inferior performance. In \textbf{S3} and \textbf{S4}, which simulate simultaneous change-points, all competing methods perform substantially worse than NetCP, with large negative Bayes factors and tight intervals below zero. For \textbf{S5}, where change-points are independent across series, G-CCP achieves the highest Bayes factor, slightly outperforming Yao, although both perform better than NetCP in this setting. These results further support the advantage of NetCP in structured or temporally dependent change-point settings, while G-CCP and Yao remain competitive in less structured scenarios.

\subsection{Seismology Data} \label{sec:seismic}
We first consider the application of our model to seismology data. We apply the NetCP model to the detection of microearthquakes using data from a network of seismic sensors. Specifically, we use recordings from the High Resolution Seismic Network (HSRN), maintained by the UC Berkeley Seismology Laboratory \parencite[]{NCEDC2014}. This network comprises geophone sensors positioned around the town of Parkfield, California, and is designed to monitor seismic activity along the San Andreas fault. Similar data was previously analyzed by \textcite{Xie2019} who took a frequentist approach, focusing on estimating the relative delay between change-points in different series. For our analysis, we consider data from seven sensors over a 100 second period beginning at 22:32:10 on 28/09/2004. Each sensor samples at 20 Hz, yielding $T = 2000$ observations per series.
\par
According to the Northern California Earthquake Catalog\footnote{https://ncedc.org/ncedc/catalog-search.html} (NCEC), four microearthquakes (events with magnitude $<$ 2.0) were recorded in Parkfield during this time. Prior to our analysis, we apply a 2-16 Hz Butterworth bandpass filter to each series, as is standard procedure in the seismology literature \parencite[]{Caudron2018}. We then normalize each series to have zero mean and unit variance. For the likelihood model within each segment we use the AR model described in Equation (\ref{eq:ar_model}), setting the lag to $L = 1$ and use hyperparameters $\alpha_j = \beta_j = \delta_{j,1} = 1$ for all series $j = 1, \dots, 7$. We then ran two MCMC chains in parallel for 20,000 iterations using the particle Gibbs scheme with 200 particles, discarding the first 2000 iterations of each chain as burn-in.
\begin{figure}[H]
    \centering
    \begin{subfigure}[t]{0.46\textwidth}
        \centering
        \includegraphics[width=1\linewidth]{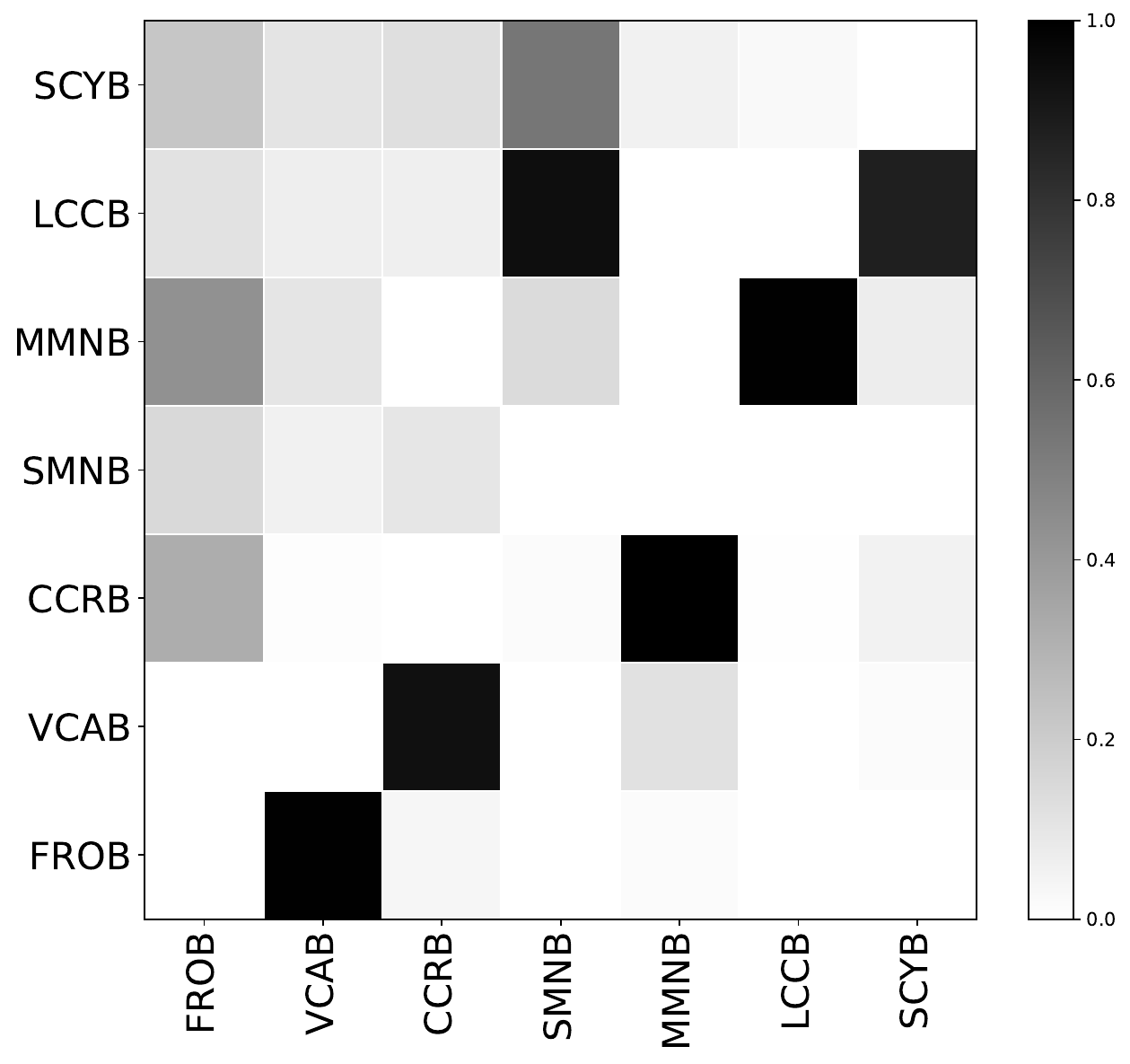} 
        \caption{}
        \label{fig:seismic_adj}
    \end{subfigure}
    \hspace{1.5em}
    \begin{subfigure}[t]{0.485\textwidth}
        \centering
        \includegraphics[width=1\linewidth]{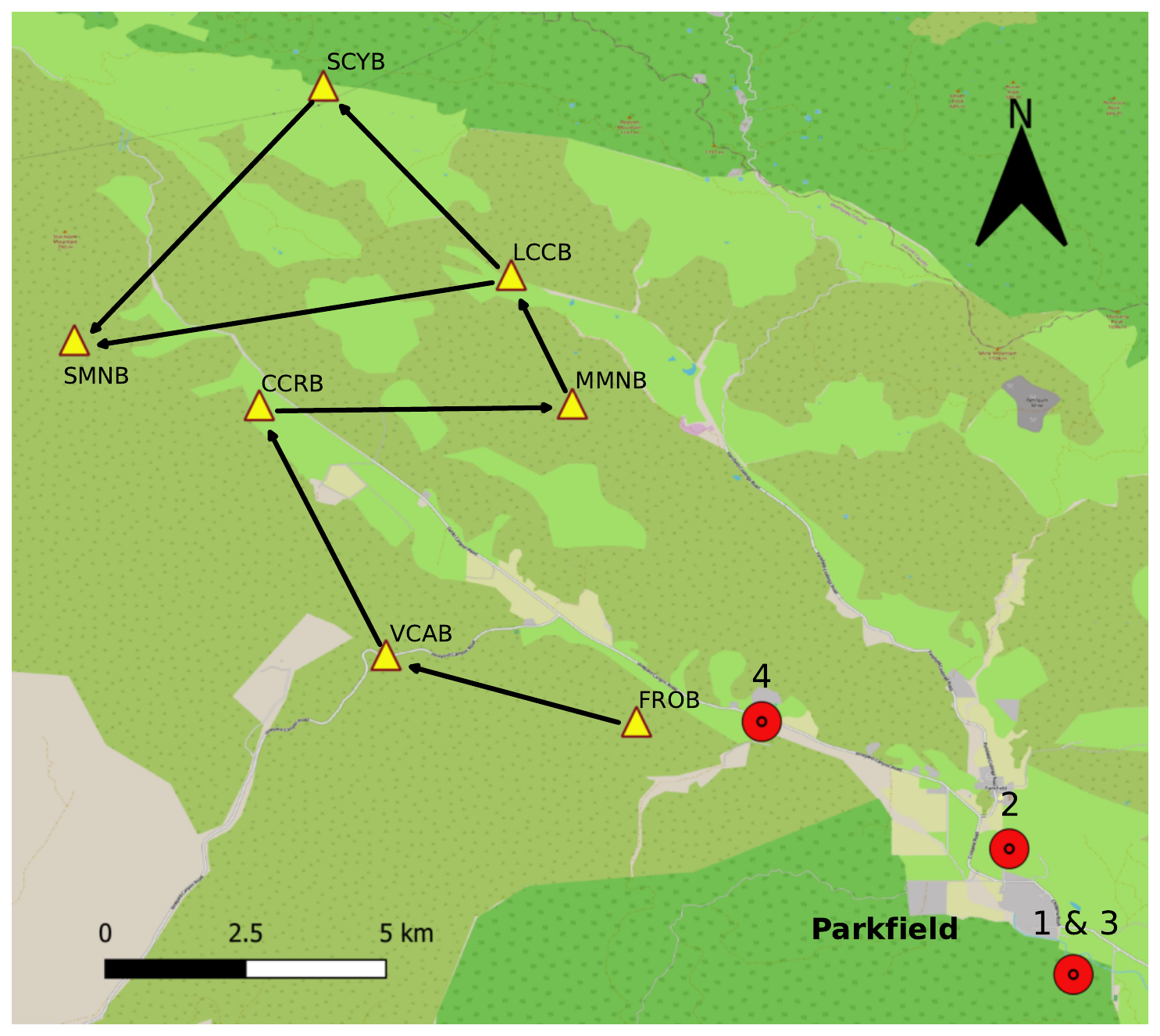} 
        \caption{}
        \label{fig:seismic_map_arrows}
    \end{subfigure}
    \caption{Network analysis of seismic sensor data: (a) Heatmap displaying estimated inclusion probabilities $Pr(A_{i, j} = 1|\boldsymbol{y})$. (b) Locations of the seven seismic sensors (yellow triangles) and the epicenters of the microearthquakes (red circles) labeled according to their order of occurrence. Arrows correspond to edges obtained after thresholding $Pr(A_{i, j} = 1|\boldsymbol{y})$ at 50\%.}
    \label{fig:seismic_map}
\end{figure}

Figure \ref{fig:seismic_map} displays the posterior edge inclusion probabilities between sensors alongside the locations of the seven seismic sensors and the microearthquake epicenters. Arrows indicate edges inferred by NetCP after thresholding the inclusion probabilities at 50\%. The resulting edges suggest that sensors nearer to Parkfield tend to lead the ones which are further away, with the sensor FROB likely to experience change-points first and SMNB likely to experience them last.
\par
\begin{figure}[H]
    \centering
    \begin{subfigure}[t]{\textwidth}
        \centering
        \includegraphics[width=1\linewidth]{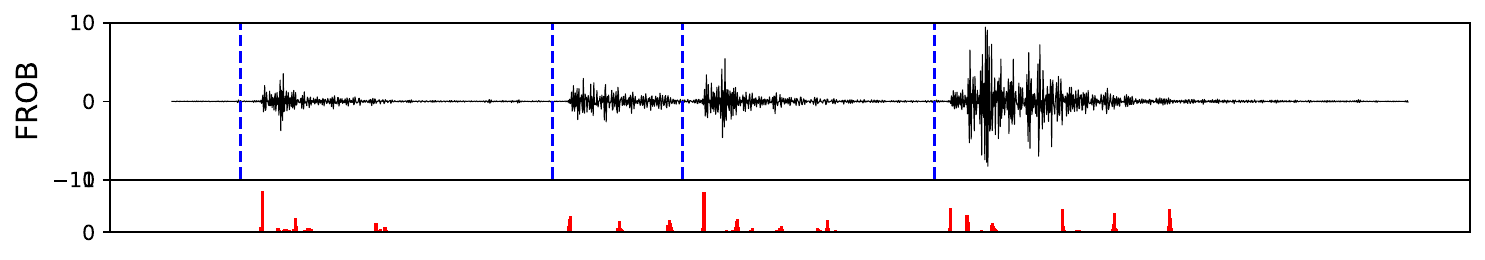} 
    \end{subfigure}
    \vfill 
    \begin{subfigure}[t]{\textwidth}
        \centering
        \includegraphics[width=1\linewidth]{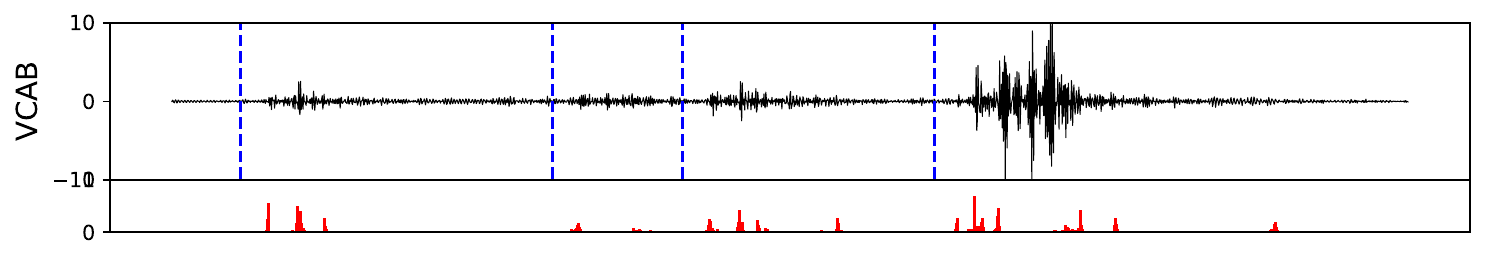} 
    \end{subfigure}
    \vfill 
    \begin{subfigure}[t]{\textwidth}
        \centering
        \includegraphics[width=1\linewidth]{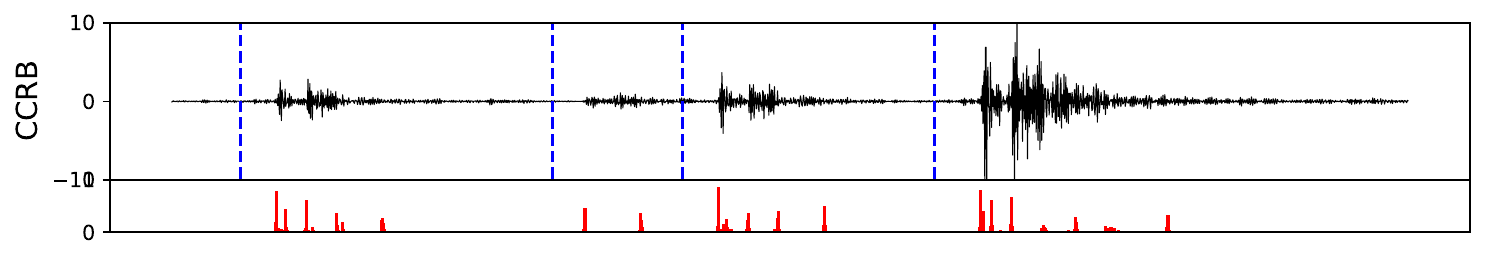} 
    \end{subfigure}
    \vfill 
    \begin{subfigure}[t]{\textwidth}
        \centering
        \includegraphics[width=1\linewidth]{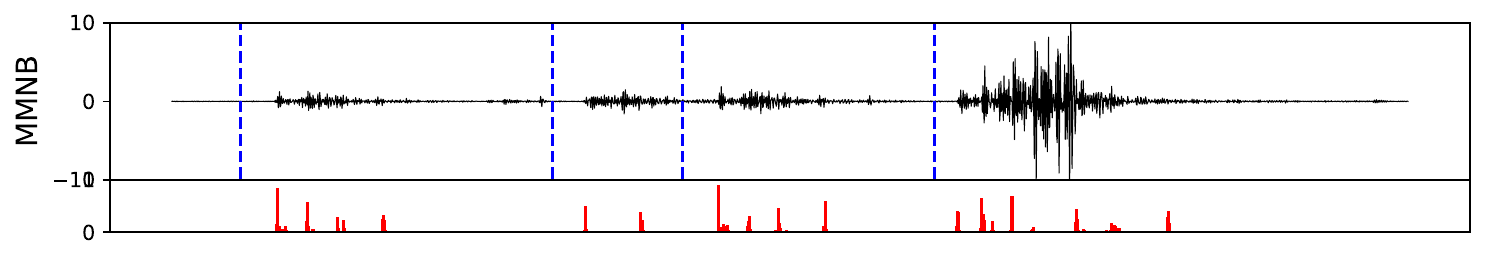} 
    \end{subfigure}
    \vfill 
    \begin{subfigure}[t]{\textwidth}
        \centering
        \includegraphics[width=1\linewidth]{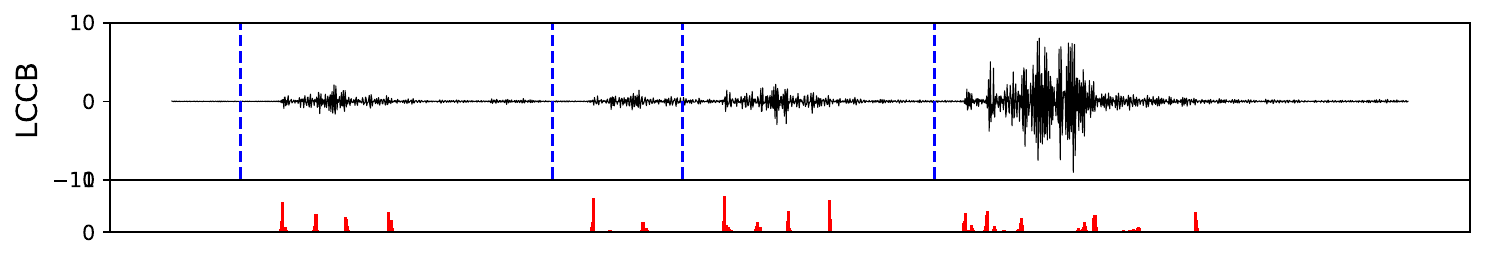} 
    \end{subfigure}
    \vfill 
    \begin{subfigure}[t]{\textwidth}
        \centering
        \includegraphics[width=1\linewidth]{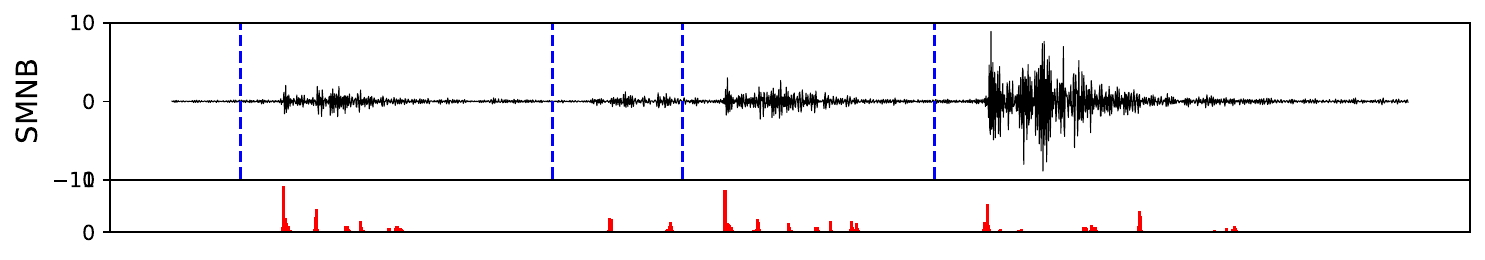} 
    \end{subfigure}
    \vfill 
    \begin{subfigure}[t]{\textwidth}
        \centering
        \includegraphics[width=1\linewidth]{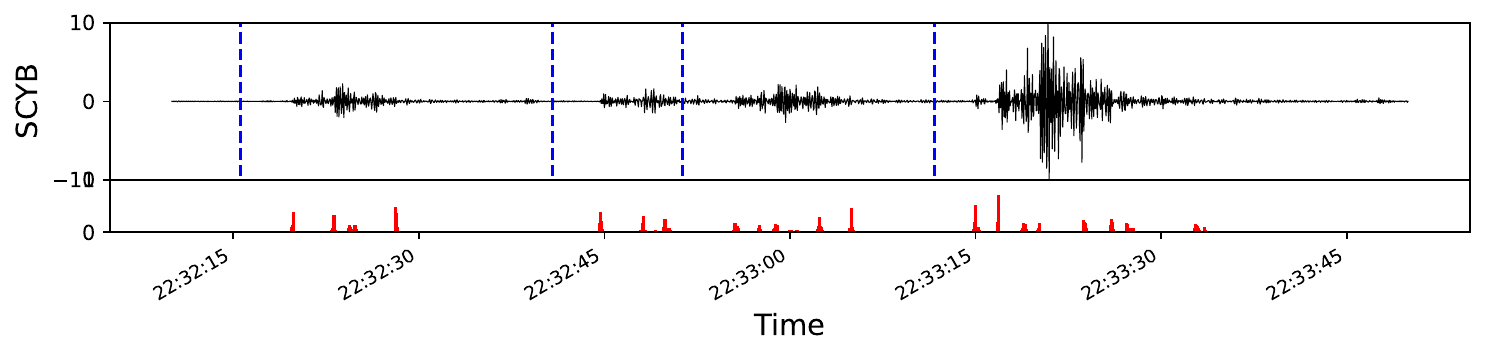} 
    \end{subfigure}
    
    \caption{Processed signals (top) and estimated posterior change-point probabilities in red (bottom) under the proposed model for the seven geophone recording stations. Blue dashed lines indicate the recorded start times for the four microearthquakes.}
    \label{fig:seis_ts}
\end{figure}

Figure \ref{fig:seis_ts} displays the processed data (top) and posterior change-point probabilities (bottom) for each seismic sensor. According to the NCEC, the first microearthquake occurs at 22:32:15.6 and its epicenter is 9km from the closest sensor - FROB, see Figure \ref{fig:seismic_map_arrows}. Our NetCP model detects the first change-point at FROB with a high probability ($\approx 0.78$) at 22:32:17.35, a delay of 1.75 seconds. It is known\footnote{See for example: www.bgs.ac.uk/discovering-geology/earth-hazards/earthquakes/how-are-earthquakes-detected/} that P-waves travel fastest in the earths crust, at around 5-7 km/s. Thus, the expected time taken for the P-waves to reach FROB is between 1.29 - 1.8 seconds, consistent with the time of the first change-point under our model. Similarly, the furthest sensor, SMNB, is located 20.8km from the first epicenter and our model detects the first change-point here at 22:32:19.05, a delay of 3.45 seconds. Again, this is consistent with the expected time of arrival for P-waves (between 2.97 - 4.16 seconds). These delays agree with the inferred posterior graph structure displayed in Figure \ref{fig:seismic_map}, which indicates that change-points propagate away from the epicenters. 

\subsection{ Electroencephalogram (EEG) Data} \label{sec:eeg}
We now apply our model to EEG signals collected from pediatric epilepsy patients at Boston Children’s Hospital. An EEG is a medical test which measures the electrical activity of the brain using electrodes placed on the scalp, and is a key tool for diagnosing and monitoring epilepsy. These recordings provide important information about the onset, origin, and spread of seizures across brain regions. Change-point analysis may be used to uncover such information. For example, change-points may first appear in the signals from the temporal lobe electrodes before propagating to signals from frontal electrodes shortly after. 
\par
\begin{figure}[H]
    \centering
    \begin{subfigure}[t]{0.49\textwidth}
        \centering
        \includegraphics[width=1\linewidth]{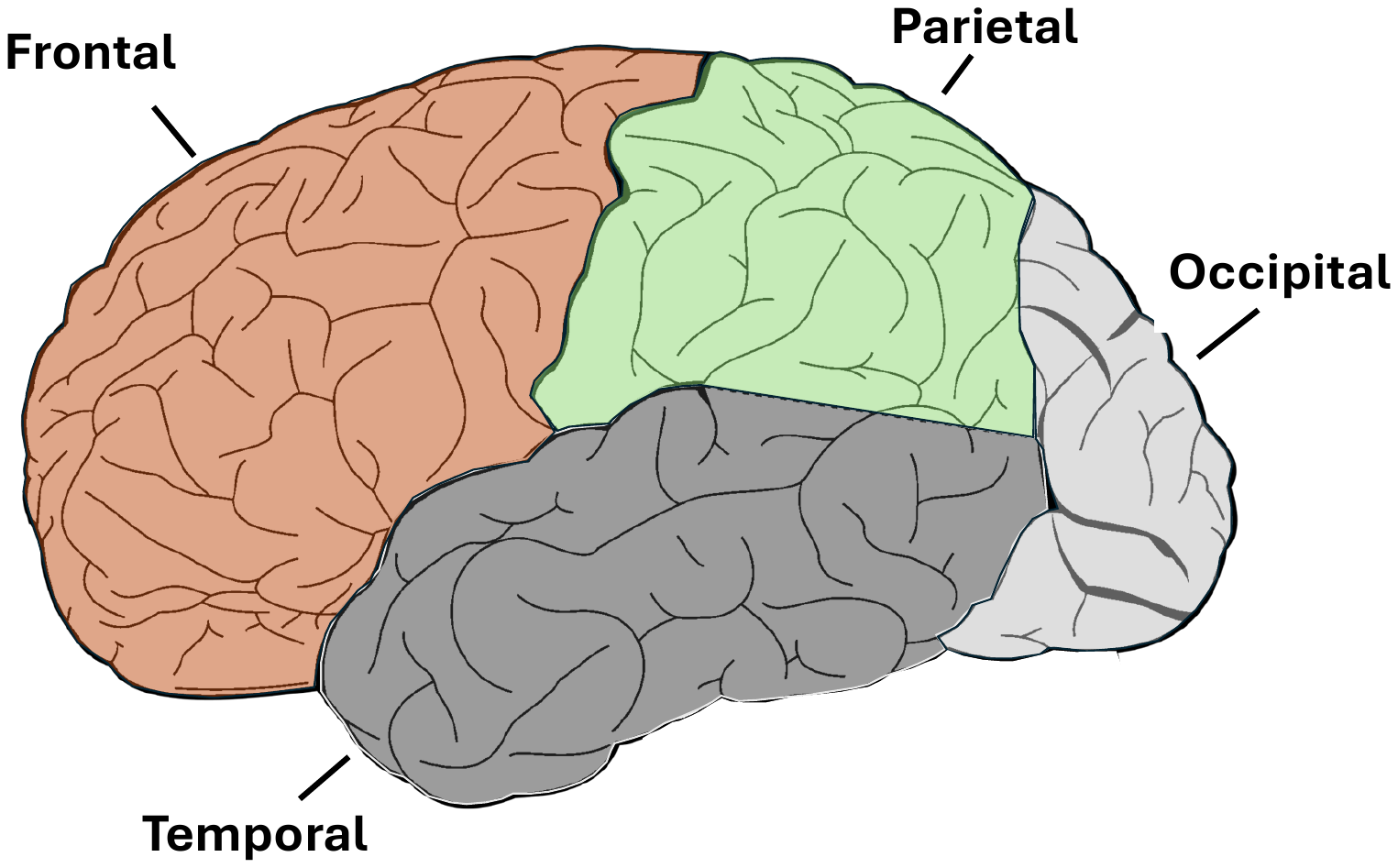} 
        \caption{}
        \label{fig:brain_side}
    \end{subfigure}
    \hfill
    \begin{subfigure}[t]{0.49\textwidth}
        \centering
        \includegraphics[width=1\linewidth]{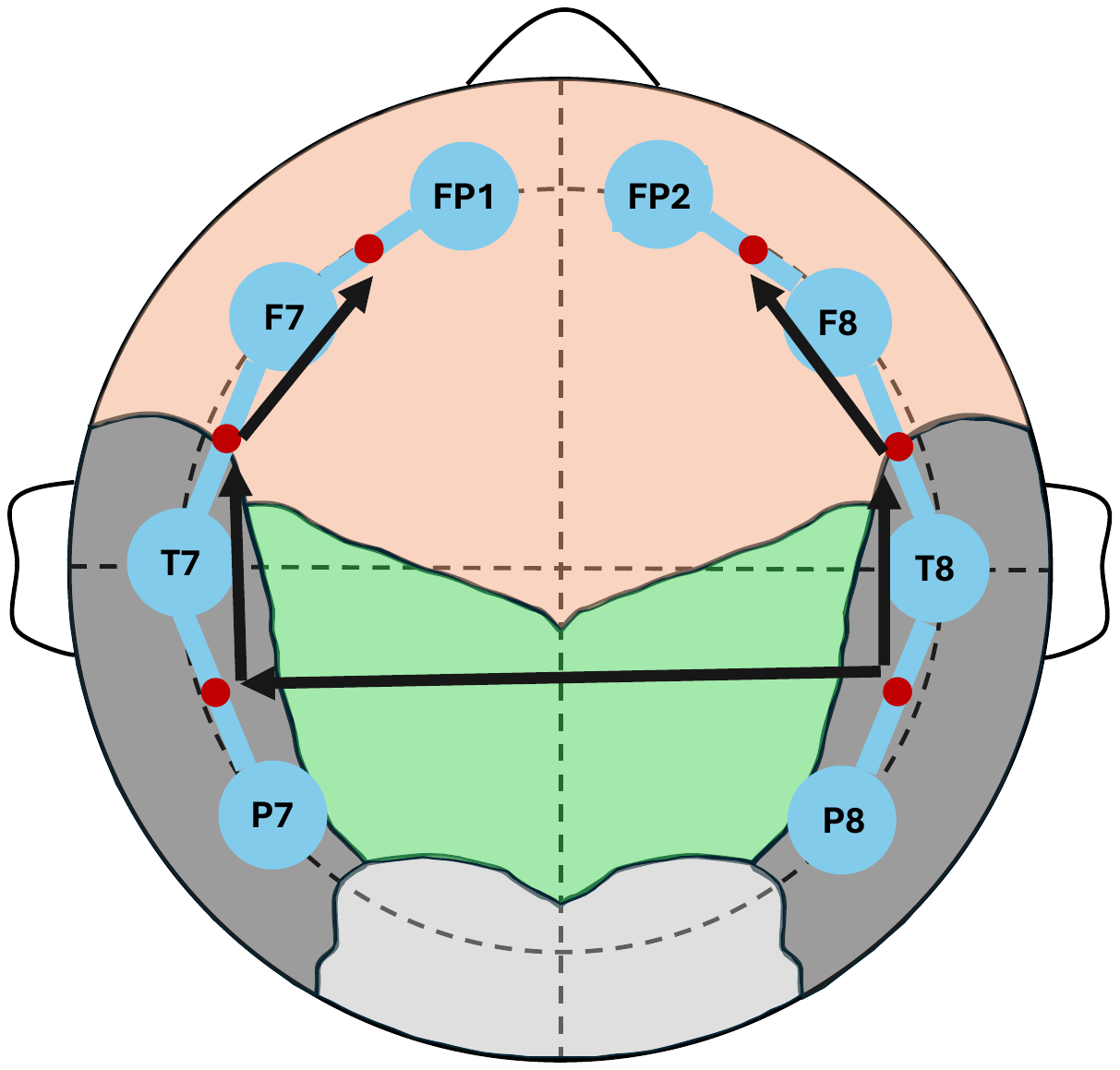} 
        \caption{}
        \label{fig:brain_top}
    \end{subfigure}
    
    \caption{EEG Electrode Positions: (a) Labeled brain lobes colored by region. (b) Top down view of electrode locations with colors corresponding to those in (a). Blue circles represent electrode locations. Blue edge with red dots indicate the bipolar montages that make up the raw signals. Black arrows indicate the edges obtained after thresholding $Pr(A_{i, j} = 1|\boldsymbol{y})$ at 50\%.}
    \label{fig:brain}
\end{figure}

The data was obtained from the CHB-MIT Scalp EEG Database \parencite[]{guttag} and was originally analyzed by \textcite{Shoeb1981}. We consider EEG data from patient CHB01, recorded during session 21. The raw signals are represented as bipolar montages, calculated as the voltage difference between pairs of adjacent electrodes in order to reduce noise. Temporal Lobe Epilepsy, characterized by seizures originating in the temporal lobes, is one of the most common forms of focal epilepsy in children. Based on this, we limit our analysis to six bipolar montages derived from eight electrodes placed in the temporal and frontal regions, see Figure \ref{fig:brain}. We focus on a 15 second segment of data from 325 to 340 seconds, during which a seizure is clinically annotated to begin at 327 seconds.
\par
Before we proceed to analyzing the EEG data, we preprocess it by applying a 1-30 Hz Butterworth bandpass filter in order to isolate the relevant frequency components \parencite[]{Chung2024}. The signals, originally sampled at 256 Hz, are then downsampled by a factor of 3. We then apply first-order differencing and normalize each series to have zero mean and unit variance. The processed dataset then consists of $d=6$ time series, each containing $T=1279$ observations. We then use the same AR likelihood model and hyperparameter settings as in Section \ref{sec:seismic} and again ran two MCMC chains for 20,000 iterations with 200 particles and a burn-in of 2000.

Figure \ref{fig:brain_side} displays the four brain lobes while Figure \ref{fig:brain_top} displays the locations of the electrodes and bipolar montages which make up the six signals. Arrows indicate the edges obtained after thresholding the posterior edge inclusion probabilities $Pr(A_{i, j} = 1 | \boldsymbol{y})$ at 50\%. The edges suggest that change-points are first observed in the right temporal lobes (electrodes T8/P8) before spreading to the left temporal lobe (electrodes T7/P7) and to the frontal lobes (electrodes FP1, FP2, F7, F8) shortly after. This suggests that the seizure originated in the region of the P8 electrode and is consistent with previous analysis by \textcite{Chung2024}, where the EEG signals were visually analyzed by two neurologists. Although this dataset does not include diagnostic information about the patient, the observed propagation pathway in Figure \ref{fig:brain_side} aligns with patterns commonly associated with Temporal Lobe Epilepsy \parencite[see][]{Jacobs2008}. Our analysis also supports the concept of epileptogenic networks, where seizures develop and propagate through localized brain networks rather than appearing as isolated events \parencite[see][]{Bartolomei2025}.
\par
\begin{figure}[H]
    \centering
    \begin{subfigure}[t]{\textwidth}
        \centering
        \includegraphics[width=1\linewidth]{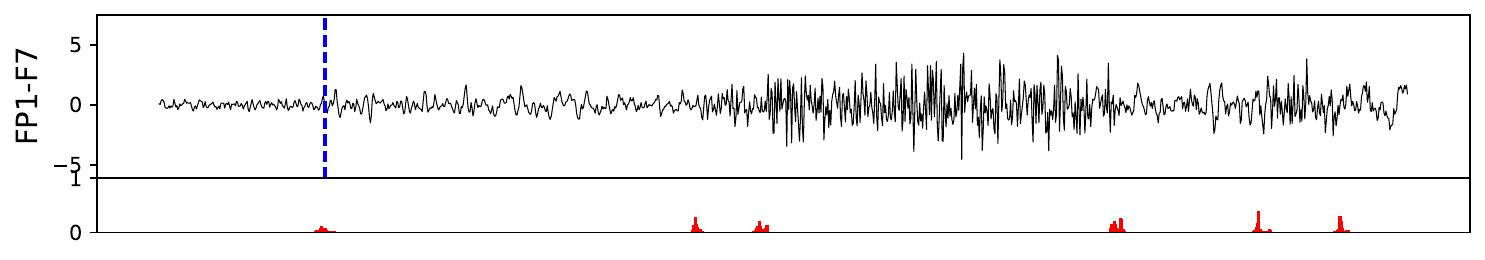} 
    \end{subfigure}
    \vfill 
    \begin{subfigure}[t]{\textwidth}
        \centering
        \includegraphics[width=1\linewidth]{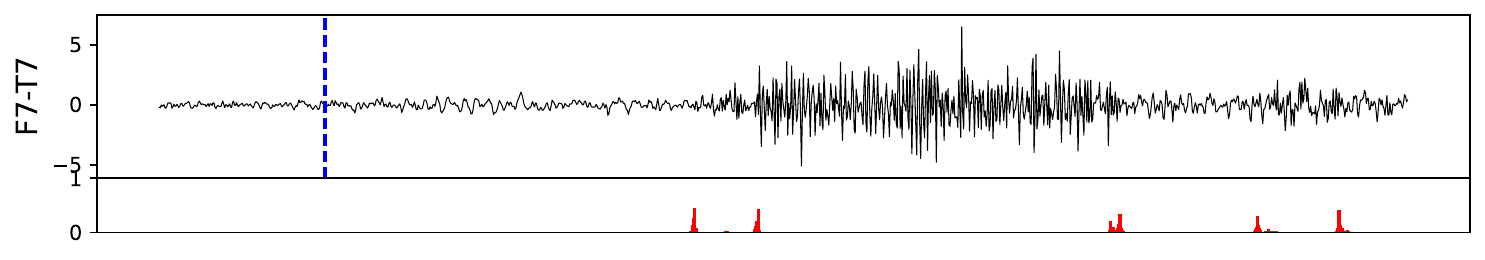} 
    \end{subfigure}
    \vfill 
    \begin{subfigure}[t]{\textwidth}
        \centering
        \includegraphics[width=1\linewidth]{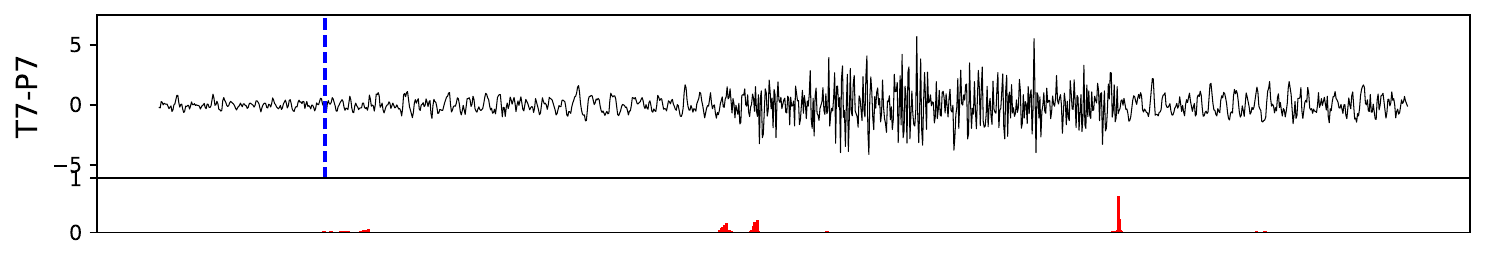} 
    \end{subfigure}
    \vfill 
    \begin{subfigure}[t]{\textwidth}
        \centering
        \includegraphics[width=1\linewidth]{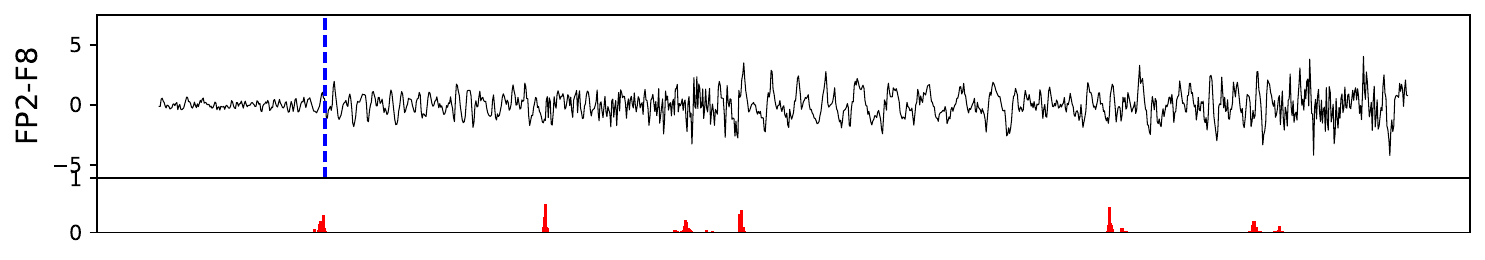} 
    \end{subfigure}
    \vfill 
    \begin{subfigure}[t]{\textwidth}
        \centering
        \includegraphics[width=1\linewidth]{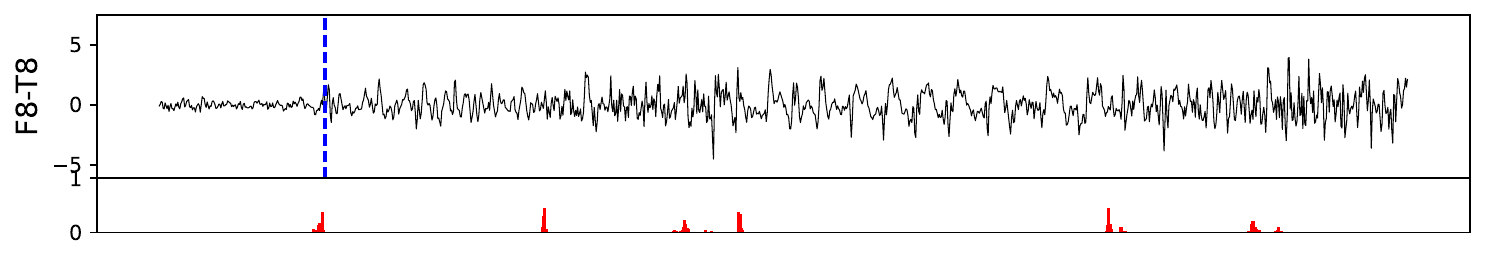} 
    \end{subfigure}
    \vfill 
    \begin{subfigure}[t]{\textwidth}
        \centering
        \includegraphics[width=1\linewidth]{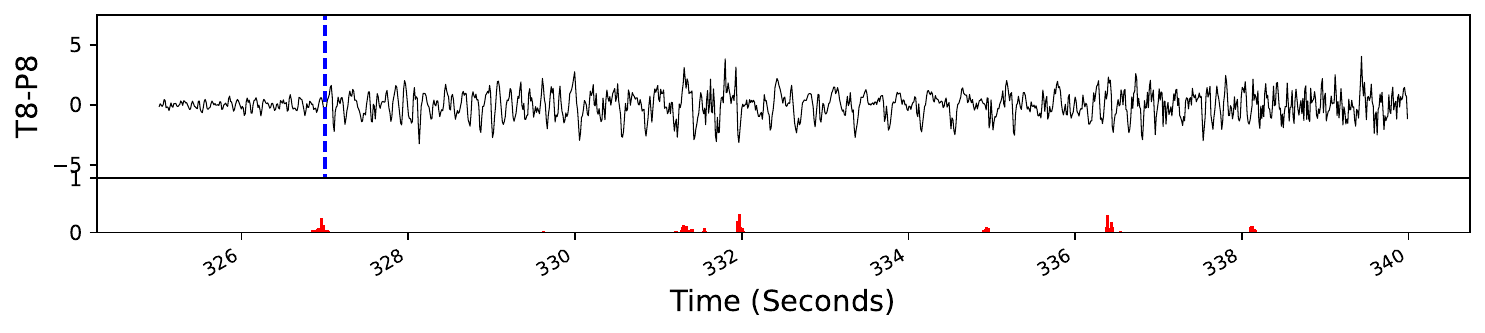} 
    \end{subfigure}
    
    \caption{Processed EEG signals (top) and estimated posterior change-point probabilities, in red, (bottom) under the proposed model for the six bipolar montages. Blue dashed lines indicate the clinically annotated seizure onset time.}
    \label{fig:eeg_ts}
\end{figure}

Figure \ref{fig:eeg_ts} displays the processed EEG signals (top) and posterior change-point probabilities (bottom). The first change-point appears in the T8-P8 signal, and is consistent with the clinically annotated seizure start time. Subsequent change-points can be seen in all or subsets of the signals as the seizure activity continues to evolve and propagate to other regions of the brain.

\section{Discussion} \label{sec:disc}
In this work, we introduced a Bayesian framework for detecting change-points across multiple time series, specifically tailored to settings where subsets of series exhibit lead-lag relationships in their change-point locations. These dependencies are captured through a latent directed graph, where edge weights encode both the strength and delay of influence from leading to lagging series. Posterior inference is made tractable by extending efficient particle MCMC techniques originally developed for univariate change-point models.
\par
Empirical results show that the proposed method outperforms existing approaches in scenarios with strong inter-series dependence in change-point structure. Moreover, the model successfully recovers interpretable latent network structures in real-world seismological and EEG datasets, providing a principled and automated way to detect and quantify lead-lag behavior previously observed in these domains.
\par
Although the MCMC algorithm employed performs well for the problems considered, scalability remains a challenge. The discrete and multimodal nature of the posterior poses difficulties for efficient sampling, and the computational cost scales linearly with the number of series, as a separate particle filter must be run for each. This highlights the need for more scalable inference methods in high-dimensional settings. 
\par
The proposed model may also be extended in several ways. A central assumption is that observations within a segment are independent of those in other segments. However, in some practical settings, this assumption can be restrictive, for example, when certain data characteristics persist or evolve gradually across segment boundaries. An extension to incorporate Markov switching dynamics, where segment-specific parameters can transition between a set of recurring states, could provide a more flexible framework for modeling data that exhibit distinct and recurring \emph{regimes}. Finally, in this paper we have assumed a fixed latent directed graph. For our future work we plan to relax this assumption and allow the graph to change at certain times, reflecting a shift in the lead-lag relationships. 

\section*{Acknowledgments}
Carson McKee was supported by the Heilbronn Institute for Mathematical Research. The simulation study was aided by the King's Computational Research, Engineering and Technology Environment (CREATE) computing cluster.

\section*{Appendix A: Proofs}
\subsection*{Proof of Proposition 1}
\noindent
According to Equation \eqref{eq:pjt_homo}, for any time $t > 1$, we have $W_{0, j}q_{0, j} / Z_j \leq p_{j}(\bx_{t-1}) < 1$ and $0 < (1-p_{j}(\bx_{t-1})) \leq 1 - W_{0, j}q_{0, j}/Z_j$. Hence the chain is irreducible and positive recurrent since there is always positive probability of the chain resetting the run length to 1 or incrementing. The chain is also clearly aperiodic. Convergence to stationarity then follows using standard results from ergodic theory, e.g. Theorem 13.0.1 of \textcite{Meyn2009}.

\subsection*{Proof of Proposition 2:}
Let $t>1$. First note that since $A_{1, 1} = A_{2, 1} = 0$, then we have $Pr(X_{1, t} = 1|\bX_{t-1}=\bx_{t-1}) = q_{0, 1}$ and $Pr(X_{1, t} = x_{1, t-1}+1|\bX_{t-1}=\bx_{t-1}) = 1-q_{0, 1}$. It is then easy to verify that
\begin{equation*}
    Pr(X_{1, t} = k) = \begin{cases}
        q_{0, 1}(1-q_{0, 1})^{k-1}, \;\;\; \text{for } k=1, ..., t-1 \\
        (1-q_{0, 1})^{k-1}, \;\;\;\;\;\;\;\;\; \text{for } k=t;
    \end{cases}
\end{equation*}
Now, for $t > 1$ and $Z_2 = W_{0, 2} + W_{1,2}$:
\begin{eqnarray*}
    Pr(U_{2, t} = 1) &=& \sum_{k=1}^{t-1}Pr(X_{2, t}=1|X_{1, t-1}=k)Pr(X_{1, t-1}=k) \\
    &=& \sum_{k=1}^{t-1} Z_2^{-1}\left[W_{0, 2}q_{0, 2} + W_{1, 2}g_{1, 2}(k)\boldsymbol{1}(k < t-1)\right]Pr(X_{1, t-1} = k) \\
    &=& Z_2^{-1}\left[W_{0, 2}q_{0, 2} + W_{1, 2}\sum_{k=1}^{t-1}g_{1, 2}(k)\boldsymbol{1}(k < t-1)Pr(X_{1, t-1} = k)\right] \\ 
    &=& Z_2^{-1}\left[W_{0, 2}q_{0, 2} + W_{1, 2}\lambda(t-1, q_{0, 1}, q_{1, 2})\right]
\end{eqnarray*}
where 
\begin{eqnarray*}
    \lambda(t, q_{0, 1}, q_{1, 2}) &=& \sum_{k=1}^t g_{1, 2}(k)\boldsymbol{1}(k<t)Pr(X_{1, t} = k) \\
    &=& \sum_{k=1}^{t-1} g_{1, 2}(k)Pr(X_{1, t} = k) \\
    &=& \sum_{k=1}^{t-1} q_{1, 2}(1-q_{1, 2})^{k-1} q_{0, 1}(1-q_{0, 1})^{k-1} \\
    &=& q_{0, 1}q_{1, 2} \sum_{k=0}^{t-2} [(1-q_{0, 1})(1-q_{1, 2})]^k \\
    &=& q_{0, 1}q_{1, 2}\left[\frac{1 - [(1-q_{0,1})(1-q_{1, 2})]^{t-1}}{1 - (1-q_{0, 1})(1-q_{1, 2})}\right]
\end{eqnarray*}
as required.
\subsection*{Proof of Proposition 3:}
Let $t > 1$ and $h\geq1$. We compute 
\begin{eqnarray*}
    \mathbb{E}(U_{1, t}U_{2, t+h}) &=& Pr(X_{1, t} = 1, X_{2, t+h}=1) \\
    &=& Pr(X_{1, t} = 1)Pr(X_{2, t+h} = 1 | X_{1, t} = 1) \\
    &=& q_{0, 1}\sum_{k=1}^{t+h-1}Pr(X_{2, t+h} = 1 | X_{1, t+h-1} = k, X_{1, t} = 1)Pr(X_{1, t+h-1}=k|X_{1, t}=1) \\
    &=& q_{0, 1}\sum_{k=1}^{t+h-1}Pr(X_{2, t+h} = 1 | X_{1, t+h-1} = k)Pr(X_{1, t+h-1}=k|X_{1, t}=1)
\end{eqnarray*}
Since the change-points in series one follow an independent Bernoulli process, we have $Pr(X_{1, t+h-1} = k|X_{1, t} = 1) = Pr(X_{1, h} = k)$ for $k \in \{1, ..., h\}$. We also have that $\boldsymbol{1}(X_{1, t+h-1} < t+h-1) = 1$. Therefore, 
\begin{eqnarray*}
    \mathbb{E}(U_{1, t}U_{2, t+h}) &=& q_{0, 1}\sum_{k=1}^{h}Z_2^{-1}[W_{0, 2}q_{0, 2}+W_{1, 2}g_{1, 2}(k)]Pr(X_{1,h} = k) \\ 
    &=& q_{0, 1}Z_2^{-1}[W_{0, 2}q_{0, 2}+W_{1, 2}\sum_{k=1}^{h}g_{1, 2}(k)Pr(X_{1, h} = k)]
\end{eqnarray*}
Now substituting $g_{1, 2}(k) = q_{1, 2}(1-q_{1, 2})^{k-1}$ and following similar steps as in Proposition 2, we obtain
\begin{eqnarray*}
    \mathbb{E}(U_{1, t}U_{2, t+h}) = q_{0, 1}Z_2^{-1}[W_{0, 2}q_{0, 2}+W_{1, 2}\lambda^*(h, q_{0, 1}, q_{1,2})]
\end{eqnarray*}
where 
\begin{equation*}
    \lambda^*(t, q_{0, 1}, q_{1, 2}) = \lambda(t, q_{0, 1}, q_{1, 2}) + q_{12}[(1-q_{0, 1})(1-q_{1, 2})]^{t-1}
\end{equation*}
and $\lambda(t, q_{0, 1}, q_{1, 2})$ is given above in Proposition 2. The result then follows combining this with $\mathbb{E}(U_{1, t}) = q_{0, 1}$ and $\mathbb{E}(U_{2, t+h}) = Pr(U_{2, t+h} = 1)$ given in Proposition 2.

\section*{Appendix B: Detailed MCMC Steps}

\subsection*{Sampling Hidden States $\bx$}
Sampling the hidden states in series $j$, $\bx_{j, 0:T}$ using Algorithm \ref{alg:particle_gibbs} requires additional algorithms to perform resampling and backward sampling. Since we are performing conditional particle filtering within a particle Gibbs framework, the previously sampled hidden states for series $j$ are guaranteed to survive all resampling steps when filtering. We therefore use the conditional SOR Algorithm described by Algorithm \ref{alg:cond_opt_resampling}. All of the below algorithms can be found in \parencite[]{Whiteley2011, Fearnhead2007}. 

\begin{algorithm}[H]
\caption{Conditional Stratified Optimal Resampling}
\label{alg:cond_opt_resampling}
\begin{algorithmic}[1]
\Require $(\mathcal{S}_t, \mathcal{W}_t)$ particles/weights to resample, $x_{j, t}^*$ (current hidden state), $N$ (number particles to keep).
\State Find $\kappa$ such that $\sum_{s \in \mathcal{S}_t} \text{min}(w_t(s)/\kappa, 1) = N$.
\State For each $s \in \mathcal{S}_t$, if $w_t(s) > \kappa$, place particle $s$ into set $A$. Otherwise place $s$ into set $B$.
\State Let $(\mathcal{S}_t^A, \mathcal{W}_t^A)$ and $(\mathcal{S}_t^B, \mathcal{W}_t^B)$ be the particles \& weights for sets $A$ and $B$ respectively. 
\State Assume there are $K$ particles in $\mathcal{S}_t^A$.
\If{$x_{j, t}^* \in \mathcal{S}_t^B$}
    \State Sample $N-K$ particles from $(\mathcal{S}_t^B, \mathcal{W}_t^B)$ using the Conditional Stratified Resampling Algorithm (conditional on $x_{j, t}^*$).
\Else 
    \State Sample $N-K$ particles from $(\mathcal{S}_t^B, \mathcal{W}_t^B)$ using the Stratified Resampling Algorithm.
\EndIf
\State The surviving particles then consist of those from $\mathcal{S}_t^A$ (with their original weights) and those that survived resampling from $\mathcal{S}_t^B$, each assigned weight $\kappa$.
\end{algorithmic}
\end{algorithm}
Resampling the particles of set $B$ in steps 6 and 8 requires either the conditional stratified resampling or stratified resampling algorithms. These are described by Algorithm \ref{alg:cond_strat_resampling} and Algorithm \ref{alg:strat_resampling}. 
\begin{algorithm}[H]
\caption{Conditional Stratified Resampling}
\label{alg:cond_strat_resampling}
\begin{algorithmic}[1]
\Require $(\mathcal{S}_t^B, \mathcal{W}_t^B)$ particles/weights to resample, $x_{j, t}^*$ (current hidden state), $N-K$ (number particles to keep).
\State Normalize the weights, $\mathcal{W}_t^B$, such that $\sum_{s\in\mathcal{S}_t^B}w_t^B(s) = 1$.
\State Construct the distribution function:
\[
    Q_t(s) = \sum_{\{s^{'} \in \mathcal{S}_t : s^{'} \leq s\}} w_{t}(s^{'})
\]
\State Sample $V^*$ uniformly on $[Q_t(x_{j, t}^*-1), Q_t(x_{j, t}^*)]$, set $V_1 = V^* - \frac{\lfloor (N-K)V^* \rfloor}{N-K}$ and $V_p=V_{p-1} + 1/(N-K)$ for $p = 2, ..., N-K$.
\State A particle $s \in \mathcal{S}_t^B$ survives resampling if there exists a $V_p$ s.t. $Q_t(s-1) \leq V_p \leq Q_t(s)$.
\end{algorithmic}
\end{algorithm}

\begin{algorithm}[H]
\caption{Stratified Resampling}
\label{alg:strat_resampling}
\begin{algorithmic}[1]
\Require $(\mathcal{S}^B_t, \mathcal{W}^B_t)$ particles/weights to resample, $N-K$ (number particles to resample).
\State Normalize the weights, $\mathcal{W}_t^B$, such that $\sum_{s\in\mathcal{S}_t^B}w_t^B(s) = 1$.
\State Construct the cumulative distribution function:
\[
    Q_t(s) = \sum_{\{s^{'} \in \mathcal{S}_t^B : s^{'} \leq s\}} w_{t}^B(s^{'})
\]
\State Sample $V_1$ uniformly on $[0, 1/(N-K)]$ and set $V_p=V_{p-1} + 1/(N-K)$ for $p = 2, ..., N-K$.
\State A particle $s \in \mathcal{S}_t^B$ survives resampling if there exists a $V_p$ s.t. $Q_t(s-1) \leq V_p \leq Q_t(s)$.
\end{algorithmic}
\end{algorithm}

After generating the particle approximations for each time $t=1, ..., T$, we generate a new sample $\bx_{j, 0:T}$ using backward sampling given in Algorithm \ref{alg:backward_sampling}.

\begin{algorithm}[H]
\caption{Backward Sampling}
\label{alg:backward_sampling}
\begin{algorithmic}[1]
\Require $\{(\mathcal{S}_t, \mathcal{W}_t)\}_{t=1, ..., T}$ particle/weight sets after performing particle filtering on series $j$, $\bx^*$ (current matrix of hidden states).
\State Sample $x_{j, T}$ from the distribution on $\mathcal{S}_T$ defined by the weights $\mathcal{W}_T$. 
\For{$t=T-1, ..., 1$}
    \State Sample $x_{j, t}$ from the distribution on $\mathcal{S}_t$ proportional to $w_t(x_{j, t})P_{t+1}^{(j)}(x_{j, t+1}|x_{j, t}, \bx_{(-j), 0:T}^*)$
\EndFor
\State Return the sampled sequence $\bx_{j, 0:T} = (x_{j, t}, \cdots, x_{j, T})$.
\end{algorithmic}
\end{algorithm}

\subsection*{Sampling $\bA$}
We sample $\bA$ by Gibbs sampling each pair $(A_{i, j}, A_{j, i})$ for $i=1, ..., d, j=i+1, ..., d$. The full posterior conditional for each pair $(A_{i, j}, A_{j, i})$ is proportional to:
\begin{align} \label{eq:sample_A}
    \pi(A_{i, j} = a_{i, j}, A_{i, j} = a_{i, j} | \boldsymbol{y}, \cdots) \\ 
    \propto \;\;Pr(A_{i, j} = a_{i, j}, A_{i, j} = a_{i, j}&| \rho) Pr(\bx_{i, 0:T}|\bW_0, \bW, \bA^*, \boldsymbol{q}_0, \boldsymbol{q}) Pr(\bx_{j, 0:T}|\bW_0, \bW, \bA^*, \boldsymbol{q}_0, \boldsymbol{q}) \\ 
    = \;\;Pr(A_{i, j} = a_{i, j}, A_{i, j} = a_{i, j} &| \rho) \\
    \times \prod_{t=1}^tPr(X_{i, t} = x_{i, t}|\bX_{t-1}&=\bx_{t-1}, \bW_0, \bW, \bA^*, \boldsymbol{q}_0, \boldsymbol{q})Pr(X_{j, t} = x_{j, t}|\bX_{t-1}=\bx_{t-1}, \bW_0, \bW, \bA^*, \boldsymbol{q}_0, \boldsymbol{q})
\end{align}
defined for $(a_{i, j}, a_{j, i}) \in \{(1, 0), (0, 1), (0, 0)\}$ and $\bA^*$ is the current adjacency matrix with the $a_{i, j}$ and $a_{j, i}$ imputed at $(i, j)$ and $(j, i)$ respectively. Thus, $(A_{i, j}, A_{j, i})$ can be updated using a categorical sample.

\subsection*{Sampling $\rho$}
The full posterior conditional of $\rho$ is given by
\begin{align}
    \pi(\rho|\cdots) &\propto Pr(\bA|\rho) f(\rho) \\
    &= (\rho/2)^{n_1}(1-\rho)^{n_2}\boldsymbol{1}(0 < \rho < 0.2)
\end{align}
where $n_1 = \sum_{i=1}^d\sum_{j=i+1}^d\boldsymbol{1}(A_{i, j}=1 \cup A_{j, i}=1)$ and $n_2 = \sum_{i=1}^d\sum_{j=i+1}^d\boldsymbol{1}(A_{i, j}=0 \cap A_{j, i}=0)$. This is a $Beta(1+n_1, 1+n_2)$ distribution truncated on the interval $(0, 0.2)$ and may be sampled using rejection sampling.

\subsection*{Sampling $(\boldsymbol{W}_0, \boldsymbol{W}, \boldsymbol{q}_0, \boldsymbol{q})$}
Due to correlation between $(W_{i, j}, q_{i, j})$, we find that it is beneficial to update these jointly for $j=1, ..., d;\; i=0, ..., d$. We update $(W_{i, j}, q_{i, j})$ with a random walk Metropolis step using independent Gaussian proposals: $W_{i, j}^\dagger \sim \mathcal{N}(W_{i, j}, 0.5^2)$ and $q_{i, j}^\dagger \sim \mathcal{N}(q_{i, j}, 0.05^2)$. We then accept $(W_{i, j}^\dagger, q_{i, j}^\dagger)$ with probability
\begin{equation*}
    \frac{Pr(\bX=\bx|W_{i, j}^\dagger, q_{i, j}^\dagger, \cdots)Ga(W_{i, j}^\dagger|1, 1)I(0 < q_{i, j}^\dagger < 1)}{Pr(\bX=\bx|W_{i, j}, q_{i, j}, \cdots)Ga(W_{i, j}|1, 1)I(0 < q_{i, j} < 1)}
\end{equation*}
\noindent
In practice, we find that thinning the above Metropolis update by around 15 steps produces good mixing in the datasets considered.

\section*{Appendix C: Comparison of Gibbs Sampling Schemes}
\begin{figure}[H]
    \centering
    \includegraphics[width=\linewidth]{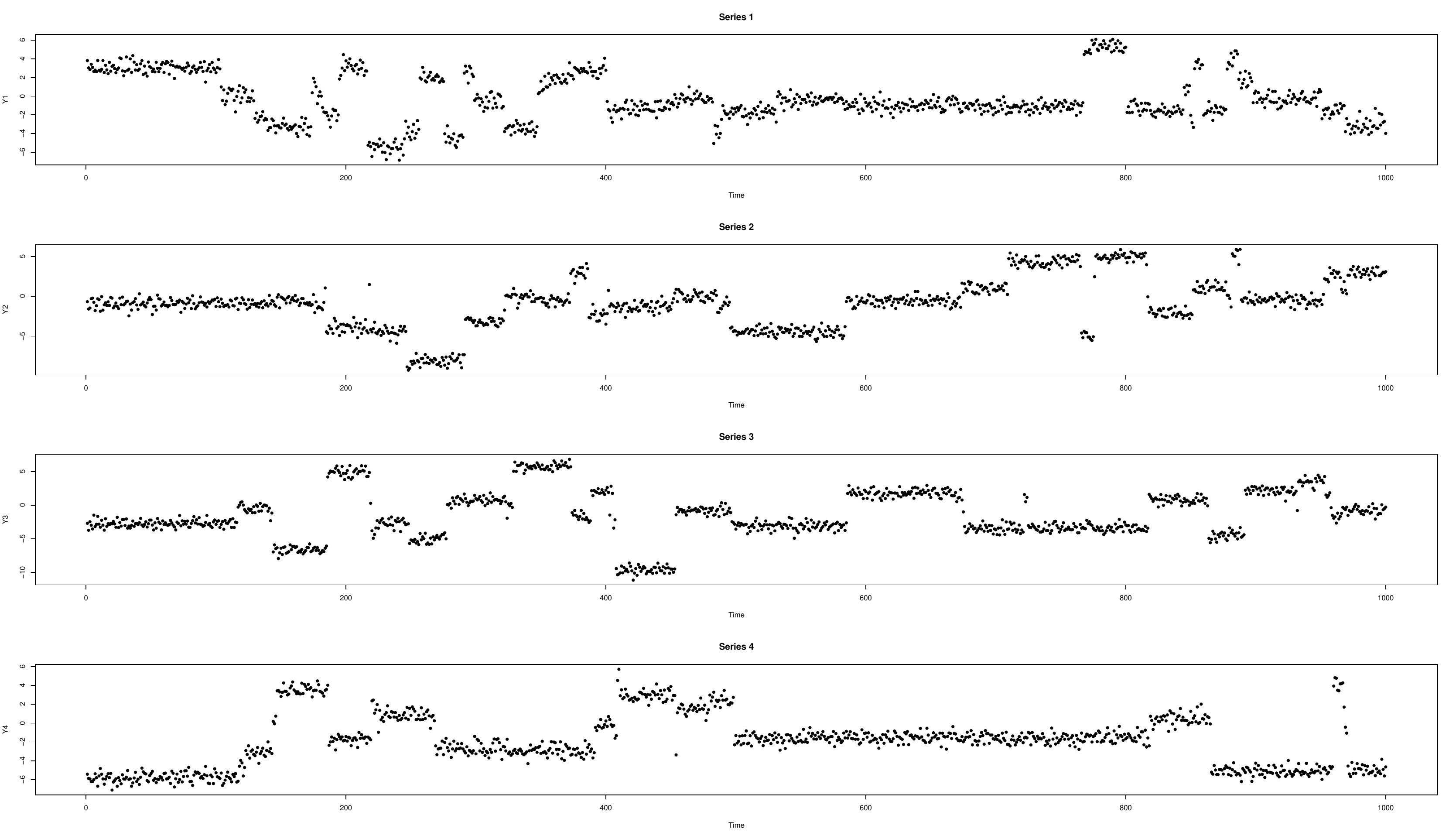}
    \caption{Simulated Times Series}
    \label{fig:sim_ts}
\end{figure}

Here we provide a comparison of the particle Gibbs scheme presented in Section \ref{sec:post} and the ``single-site'' Gibbs sampler which samples each $X_{j, t}$ one at a time from their full posterior conditionals. We simulate data with $d=4$ series and $T=1000$ time points from scenario \textbf{S1} with the Gaussian means likelihood model from from Section \ref{sec:sim_study}. The simulated data can be seen in Figure \ref{fig:sim_ts}. We then estimated the NetCP model using the same hyperparameters given in Section \ref{sec:sim_study} using (i) single-site Gibbs (SSG), (ii) particle Gibbs with 50 particles (PG50) and (iii) particle Gibbs with 100 particles (PG100). For each sampler we generated 20,000 posterior samples, discarding the first 2000 as burn-in.
\par
\begin{figure}[H]
    \centering
    \includegraphics[width=\linewidth]{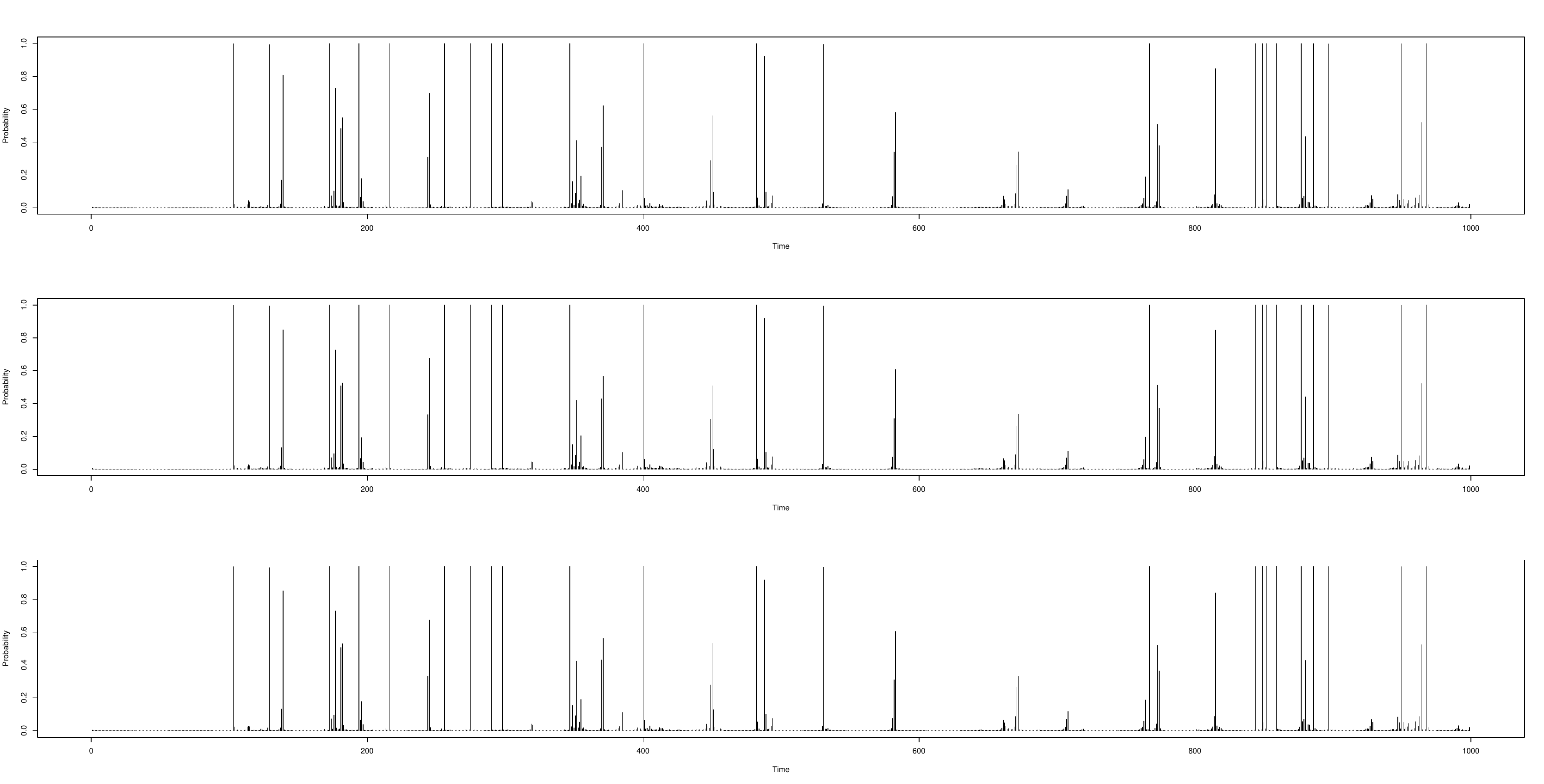}
    \caption{Estimated marginal posterior change-point probabilities for series one using SSG (top), PG50 (middle) and PG100 (bottom).}
    \label{fig:marg_probs}
\end{figure}

The estimated marginal posterior change-point probabilities for series one under each sampling method are shown in Figure \ref{fig:marg_probs}. We can see that the estimated probabilities for each method are virtually identical. In order to assess the mixing properties of the methods, we may inspect the sampled hidden states at each iteration. From Figure \ref{fig:marg_probs} we can see that in series one, time point 583 has around 0.6 probability of being included as a change-point. We can assess the mixing of $X_{1, 583}$ by inspecting the sampled values of $U_{1, 583} = \boldsymbol{1}(X_{1, 583} = 582)$. I.e. $U_{1, 583} = 1$ for a change-point and $U_{1, 583} = 0$ for no change-point. 

\begin{figure}[H]
    \centering
    \includegraphics[width=\linewidth]{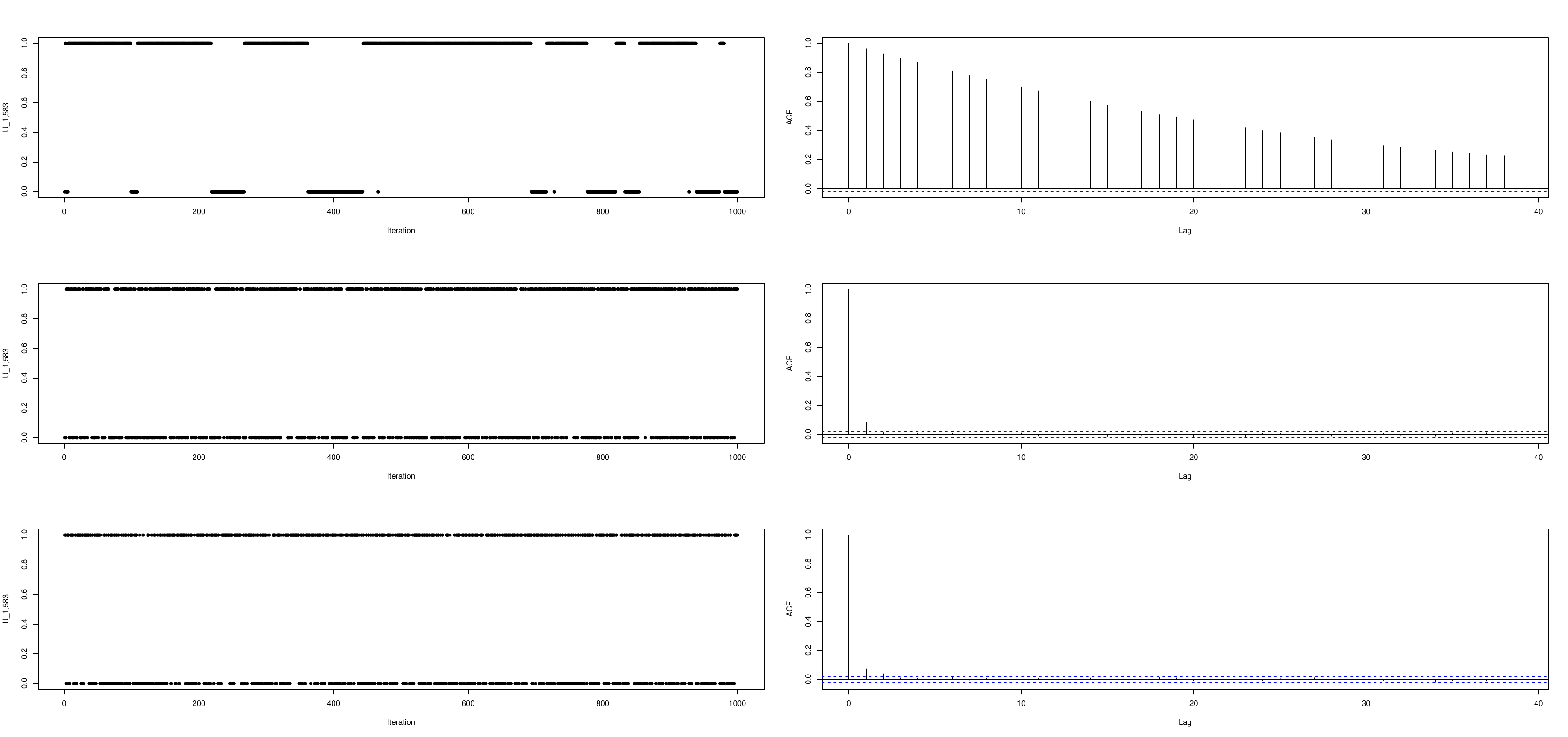}
    \caption{First 1000 samples after burn-in (left column) and estimated autocorrelation function (right column) of $U_{1, 583}$ for SSG (top row), PG50 (middle row) and PG100 (bottom row).}
    \label{fig:trace_acf}
\end{figure}

From Figure \ref{fig:trace_acf} we can see that under the single-site Gibbs sampler, $U_{1, 583}$ mixes very poorly, getting stuck in a given state for long periods of time. However, the particle Gibbs methods with 50 and 100 particles both appear to be mixing much better. This is evident in the autocorrelation plots in Figure \ref{fig:trace_acf}. Here the single-site Gibbs exhibits a slow decay in the correlation between samples while the particle Gibbs samples decorrelate almost immediately. We then estimated the integrated autocorrelation time (IAT) of each method using the R function ``IAT''. The single-site Gibbs obtains an IAT of 58.5 while the particle Gibbs schemes both obtain IATs of 1.2. This suggests that we must obtain around 49 samples from the single-site sampler in order to obtain the same amount of information as one sample from particle Gibbs samplers (when estimating $\mathbb{E}[U_{1, 583}|\boldsymbol{y}]$). Similar results can be obtained by inspecting the change-point indicators at other time points. These results suggest that relatively few particles are required to achieve good mixing, with the particle Gibbs sampler obtaining the same IAT for 50 and 100 particles.

\section*{Appendix D: Marginal Segment Likelihood for Gaussian Mean Model}
Under the Gaussian changing means likelihood model considered in Section \ref{sec:sim_study}, the likelihood for a given segment $s:t$ in series $j$ is given by 
\begin{equation*}
    f_j(\boldsymbol{y}_{j, s:t}|\theta) = \prod_{k=s+1}^t\mathcal{N}(y_{j, k}|\theta, \sigma_j^2)
\end{equation*}
and the prior on $\theta$ is given by $f_j(\theta) = \mathcal{N}(\theta|0, \gamma_j^2)$. Upon integrating over $\theta$ within each segment, we obtain the marginal segment density
\begin{equation*}
    f_j(\boldsymbol{y}_{j, s:t}) = \frac{1}{(2\pi\sigma_j^2)^{(t-s)/2}}\left(\frac{\sigma_j^2}{(t-s)\gamma_j^2 + \sigma_j^2}\right)^{1/2}\text{exp}\left(-\frac{\sum_{k=s+1}^ty_{j, k}^2}{2\sigma_j^2} + \frac{\gamma_j^2\left[\sum_{k=s+1}^ty_{j, k}\right]^2}{2\sigma_j^2\left((t-s)\gamma_j^2 + \sigma_j^2\right)}\right)
\end{equation*}
which depends on the hyperparameters $(\sigma_j^2, \gamma_j^2)$.
\vspace{1in}
\printbibliography
\end{document}